\documentclass[fleqn,usenatbib, referee]{mnras}

\usepackage{newtxtext,newtxmath}
\usepackage{subfig}
\usepackage{float}
\usepackage{footnote}
\usepackage{color,soul}

\makesavenoteenv{tabular}

\usepackage[T1]{fontenc}
\usepackage{ae,aecompl}
\usepackage{gensymb}

\usepackage{graphicx}	
\usepackage{amsmath}	
\usepackage{amssymb}	
\usepackage{multirow}

\title[MAXI J1409$-$619]{Comprehensive Analysis of the Transient X-ray Pulsar MAXI 
J1409$-$619}

\author[D\"{o}nmez et al.]
{\c{C}. K. D\"{o}nmez$^{1}$\thanks{E-mail: cagatay.donmez@metu.edu.tr},
M.M. Serim$^{1,2}$,
S. \c{C}. \.{I}nam$^{3}$\thanks{E-mail: inam@baskent.edu.tr},
\c{S}. \c{S}ahiner$^{4}$,
D. Serim$^{1}$ 
\newauthor and A. Baykal$^{1}$\thanks{E-mail: altan@astroa.physics.metu.edu.tr}
\\
$^{1}$Physics Department, Middle East Technical University, 06800 Ankara, Turkey\\
$^{2}$T\"{U}B\.{I}TAK ULAKB\.{I}M, 06510 Ankara, Turkey\\
$^{3}$Department of Electrical and Electronics Engineering, 
Ba\c{s}kent University, 06790 Ankara, Turkey\\
$^{4}$Department of Electronics and Communication Engineering, 
Beykent University, 34398 \.{I}stanbul, Turkey 
}

\date{Accepted XXX. Received YYY; in original form ZZZ}

\pubyear{2019}

\begin{document}
\label{firstpage}
\pagerange{\pageref{firstpage}--\pageref{lastpage}}
\maketitle


\begin{abstract}

We probe the properties of the transient X-ray pulsar MAXI J1409$-$619 through \textit{RXTE} and \textit{Swift} follow up observations of the outburst in 2010. We are able to phase connect the pulse arrival
times for the 25 days episode during the outburst. We suggest that either an orbital model (with $P_{{\rm{orb}}} \simeq 14.7(4)$ days) or a noise process due
to random torque fluctuations (with $S_r \approx 1.3 \times 10^{-18}$ Hz$^2$ s$^{-2}$ Hz$^{-1}$) is plausible to describe
the residuals of the timing solution. The frequency derivatives  indicate a positive torque-luminosity
correlation, that implies a temporary accretion disc formation during the outburst. We also discover several quasi-periodic oscillations (QPOs) in company with their harmonics  whose centroid frequencies decrease as the source flux decays. The variation of 
pulsed fraction and spectral power law index of the source with X-ray flux is interpreted as the sign of transition from a critical to a sub-critical accretion regime at the critical luminosity within the range of $6\times 10^{37}$ erg s$^{-1}$ to $1.2\times 10^{38}$ ergs s$^{-1}$. Using pulse-phase-resolved spectroscopy, we show that the phases with higher flux tend to have lower photon indices, indicating that the polar regions produce spectrally harder emission.

\end{abstract}





\section{Introduction}

MAXI J1409$-$619 is a transient X-ray pulsar 
which was discovered by the Gas Slit Camera (GSC), one of the detectors of the Monitor of All-sky X-ray Image (\textit{MAXI}) experiment, on 2010 October 17 with a (41$\pm$7) mCrab peak flux in 4--10 keV energy band \citep{ATel2959}. Using Neil Gehrels Swift Observatory's (\textit{Swift}) X-ray Telescope (XRT) observations, \citet{ATel2962} reported the source location as RA = 14$^{h}$ 08$^{m}$ 02.56$^{s}$, 
Dec. = $-61^\circ$ 59$^\prime$ 00.3$^{\prime\prime}$ in J2000 coordinates with an accuracy of 1.9 arcseconds. The Two Micron All Sky Survey (2MASS) star 2MASS 14080271$-$6159020 was considered as the probable infrared (IR) counterpart of the source  since it is 2.1 arcseconds away from the source location \citet{ATel2962}.
\citet{Orlandini2012} emphasised that no catalogued optical source is present at the source location (within an upper limit of 20 for R-band magnitude) and suggested that this IR companion is a highly  reddened late O/early B-type star, implying the High Mass X-ray Binary (HMXB) nature of the system. In addition, the closest source in GAIA catalogue\footnote{\url{http://gaia.ari.uni-heidelberg.de/singlesource.html}} is located 3.6 arcseconds away from the reported  X-ray position of the source \citep{ATel2962} and hence it most likely does not represent the optical counterpart of the HMXB. 
On 2010 November 30, when MAXI J1409$-$619 underwent an outburst becoming $\sim$7 times brighter 
compared to its initial observations, \textit{Swift} Burst Alert Telescope (BAT) was triggered to observe the 
source, revealing a 503$\pm$10 s periodicity  with 42\% pulsed fraction \citep{ATel3060}. Routine 
\textit{MAXI}/GSC observations also indicated that the source was brightened significantly
\citep{ATel3067}. 
Utilising the \textit{Fermi} Gamma-ray Burst Monitor (GBM) observations 
on 2010 December 2--3, the source period was refined as 506.93(5)s, 
a frequency derivative of 1.66(14)$\times$10$^{-11}$ Hz s$^{-1}$ was 
found, and the double-peaked nature of the pulse profile was revealed 
\citep{ATel3069}. The spectra obtained from the Rossi X-ray Timing Explorer (\textit{RXTE})--Proportional Counter Array (PCA) observations on December 4, 2010 with a net exposure of 8.5 ksec were 
modelled by a cutoff power law with either partially covered absorption or reflection of X-ray photons by cold electrons (see \cite{reflection2010} for a detailed discussion about the reflection model) and a narrow iron line in both cases, with no sign of cyclotron absorption feature \citep{ATel3070}. In both models, 
the photon index was found to be $\Gamma\approx1.3$ while the partially covered absorber $1.3\pm0.6\times 10^{24}$ cm$^{-2}$ is reported for the first model or the reflection scaling factor of $0.52\pm0.18$ for the latter \citep{ATel3070}. \cite{Orlandini2012} rediscovered the source from the archival \textit{BeppoSAX} observations in 2000 at its low state. From these observations, 1.8--100 keV energy spectrum was found to fit a absorbed power law model with the photon index and hydrogen column density of $0.87^{+0.29}_{-0.19}$ and $(2.8^{+3.4}_{-2.2})\times 10^{22}$ cm$^{-2}$. In addition to this spectral model, this broadband spectrum revealed the presence of cyclotron absorption feature with a fundamental energy of 44 keV leading to a magnetic field estimate of $3.8\times 10^{12}(1+z)$ Gauss, where $z$ is the gravitational redshift.

\citet{ATel3082} announced the detection of a QPO at 0.192$\pm$0.006 Hz with 
2 harmonics, using a total of 20 ks of observations on 2010 December 11. 
From archival observations of \textit{BeppoSAX}, The Advanced Satellite for Cosmology 
and Astrophysics (\textit{ASCA}), and INTErnational Gamma-Ray Astrophysics Laboratory (\textit{INTEGRAL}), it was inferred that MAXI J1409$-$619 was in a low state during the observations prior to its discovery \citep{ATel2965, Orlandini2012}.

In this paper, we present the results of comprehensive timing and spectral 
analysis of MAXI J1409$-$619, using the observations of \textit{RXTE} and \textit{Swift} observatories. 
In Section \ref{sec:obs}, we give brief information about the observations
and data reduction process. 
In Section \ref{sec:data}, we present our timing and spectral analysis. 
In Section \ref{sec:discuss}, we summarise and discuss our results.

\section{Observations and Data Reduction}
\label{sec:obs}
\subsection{\textit{RXTE}/PCA Observations}
\textit{RXTE}, launched on 1995 December 30 
and operational until 2012, was an observatory satellite dedicated 
to X-ray astronomy\footnote{The official cookbook for \textit{RXTE} data analysis is located at \url{https://heasarc.gsfc.nasa.gov/docs/xte/recipes/cook_book.html}.}. PCA instrument 
onboard \textit{RXTE} consisted of five identical 
proportional counter units (PCU), each of which was sensitive to photon energies between 2--60 keV \citep{Jahoda1996, Jahoda2006}. 

\textit{RXTE}/PCA observations of MAXI J1409$-$619 
took place between 2010 October 22 and 2011 February 5 (corresponding to the observation sets with proposal numbers 95358, 95441, 96410). A total of 50 observations are used in our analysis, amounting a total exposure of 135.0 ks. 
\textit{Standard-2} mode data are used for extracting spectra (16 s timing resolution), and 
\textit{Good Xenon} mode data are used for extracting lightcurve 
($\sim$1 $\mu$s timing resolution).

In order to process \textit{RXTE}/PCA data, good time intervals are selected. To apply background correction to spectra and lightcurve, Epoch 5C 
background models being the latest models for the observations after 2004 provided by the {\textit{RXTE} Guest Observer Facility\footnote{\url{https://heasarc.gsfc.nasa.gov/docs/xte/pca_bkg_epoch.html}}  are used. 
Lightcurve 
count rates are corrected to the total number of active PCUs and Solar System barycentric correction is applied. As it was reported to be the most reliable PCU during our observations\footnote{\url{https://heasarc.gsfc.nasa.gov/docs/xte/recipes/pca_breakdown.html}}, only PCU 2 data are used for 
spectral analysis, which leads to a total net exposure of 122.5 ks. The energy range used for spectral analysis is 3--25 keV, and 
0.5\% systematic error is applied to long and combined observations.

\subsection{\textit{Swift}/XRT Observations}
\textit{Swift} is a multi-wavelength space observatory, launched on 2004 November 20. 
\textit{Swift} is mainly used for gamma-ray burst monitoring, 
yet it also proved its effectiveness in X-ray astronomy through 
onboard 3.5-meter focal length X-ray telescope XRT operating in 0.2--10 keV energy range \citep{Gehrels2004}. Among a number of different operating modes of XRT, window timing (WT) mode   has 1.8 ms high timing resolution but keeps only one-dimensional imaging data, 
whereas photon counting (PC) mode  retains full spectroscopy and 600$\times$602 pixels imaging 
information but with only 2.5 s timing resolution \citep{Burrows2005}

32 \textit{Swift}/XRT observations of 
MAXI J1409$-$619 were made between 2010 October 20 and 2011 January 28. Of these 
observations, 13 of them were conducted in PC mode, 17 of them 
in WT mode and 2 of them utilises both modes.  Our analysis is based on the PC mode observations, each of which spans  $\sim$1--2 ks, leading to a total exposure of 19.3 ks.

Data reduction is conducted using \textsc{heasoft v6.22.1}. 
Raw \textit{Swift}/XRT data are reprocessed by the standard pipeline tool 
\textsc{xrtpipeline v0.13.4} with default screening criteria.  When the flux level of the source exceeds the pile-up limit ($\sim 0.5$ cts s$^{-1}$), pile-up corrections are performed. The source counts 
are gathered from a 20-pixel radius circle centred on the source while an inner circle of $\sim5$-pixel radius is excluded due to pile-up. Background emissions
are estimated from source-free circular regions of 50-pixel radius. 
First 30 channels of XRT data, corresponding to <0.3 keV are marked "bad" 
as XRT is not considered reliable below this energy range. The spectra are rebinned through \textsc{grppha} in such a way that 
there are at least 5 counts per bin to obtain statistically significant spectra.

\section{Data Analysis}
\label{sec:data}

\subsection{Timing Analysis}
\subsubsection{Pulse Timing, Timing Solution and Torque Noise}
\label{sec:timing}
In the timing analysis, we use 0.125 s  binned \textit{RXTE}/PCA lightcurve which is extracted using the procedure described in Section \ref{sec:obs}.
The 1-day binned lightcurve is shown in Figure \ref{fig:lcurve}. 

\begin{figure}
	\includegraphics[width=\columnwidth]{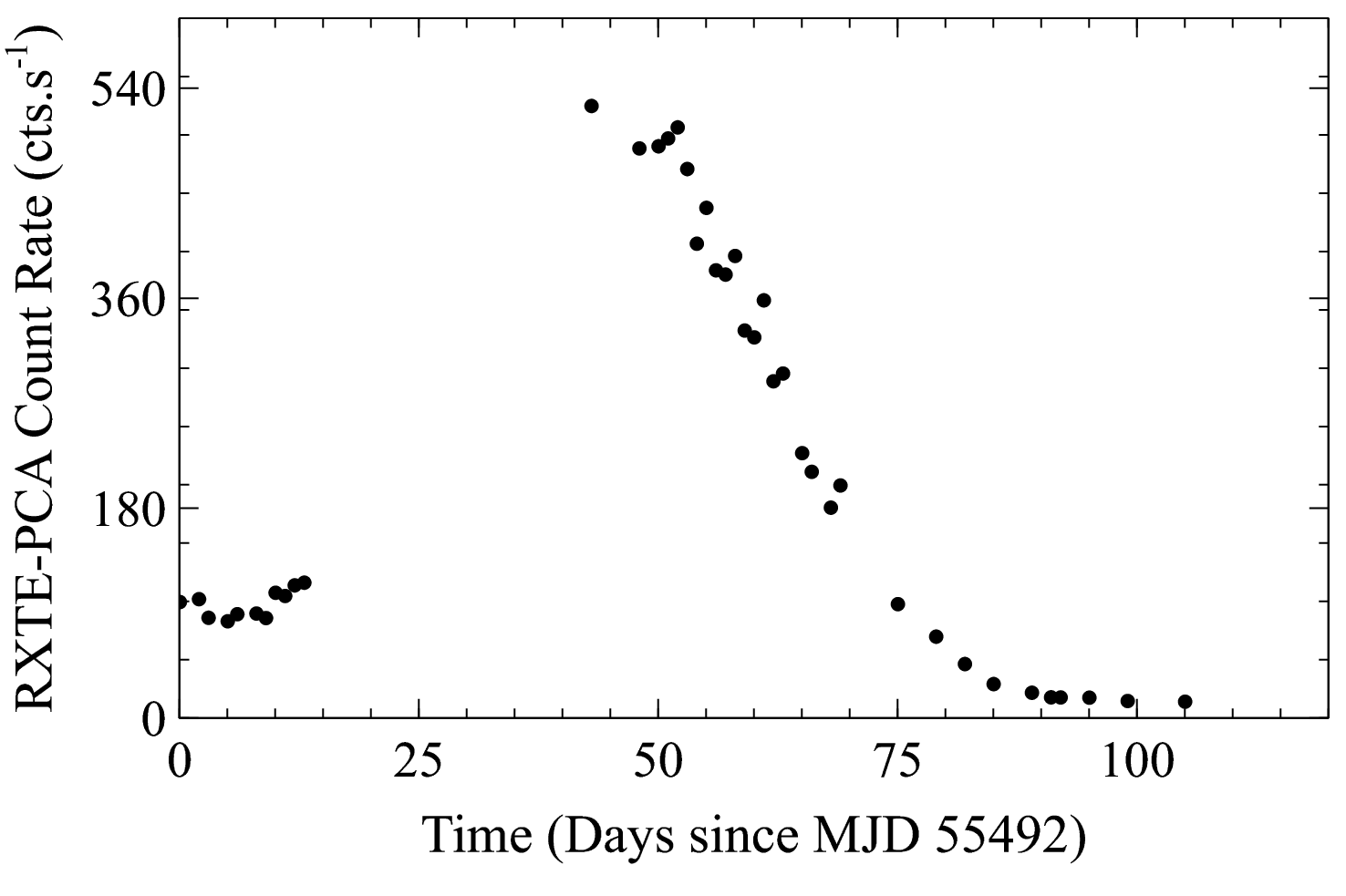}
	\caption{2--60 keV 1-day binned \textit{RXTE}/PCA lightcurve of MAXI J1409$-$619. Errors are smaller  than the data points.}
	\label{fig:lcurve}
\end{figure}

We search for the best frequencies by using statistically independent trial frequencies \citep{Leahy1983}.
Our search of best frequencies is consistent with the pulse frequencies provided by \textit{Fermi}/GBM pulsar project team\footnote{\url{https://gammaray.nsstc.nasa.gov/gbm/science/pulsars.html}}.
By folding the lightcurve at pulse frequency, we construct the pulse profiles with 20 phase bins for each observation. After performing $\chi^2$ test for all the constructed pulse profiles, the statistically strongest profile is assigned as template pulse profile.
Then, the pulse arrival times are computed by cross-correlating each individual pulse profile with the template.
In the cross-correlation analysis, we use the harmonic representation of pulses 
(\citet{DeeterBoynton1985}, see also \citet{Icdem2012} for applications).

We are able to phase connect all pulse arrival times during the outburst.
We fitted pulse arrival times to the cubic polynomial,
\begin{equation}
\delta \phi = \delta \phi_{o} + \delta \nu (t-t_{o}) 
+ \frac{1}{2}\dot{\nu}  (t-t_{o})^{2} 
+ \frac{1}{6}\ddot{\nu} (t-t_{o})^{3}
\label{polyn}
\end{equation}
where $\delta \phi$ is the pulse phase offset, $t_{{\rm{o}}}$ is the epoch for folding;
$\delta \phi_{o}$ is the residual phase offset at the time t$_{o}$; $\delta \nu$ is the
correction to the pulse frequency at 
$t_0$  with respect to the folded frequency $\nu_0$; $\dot{\nu}$ and $\ddot{\nu}$ are the first and second time derivatives of pulse frequency.

\begin{figure}
	\includegraphics[width=\columnwidth]{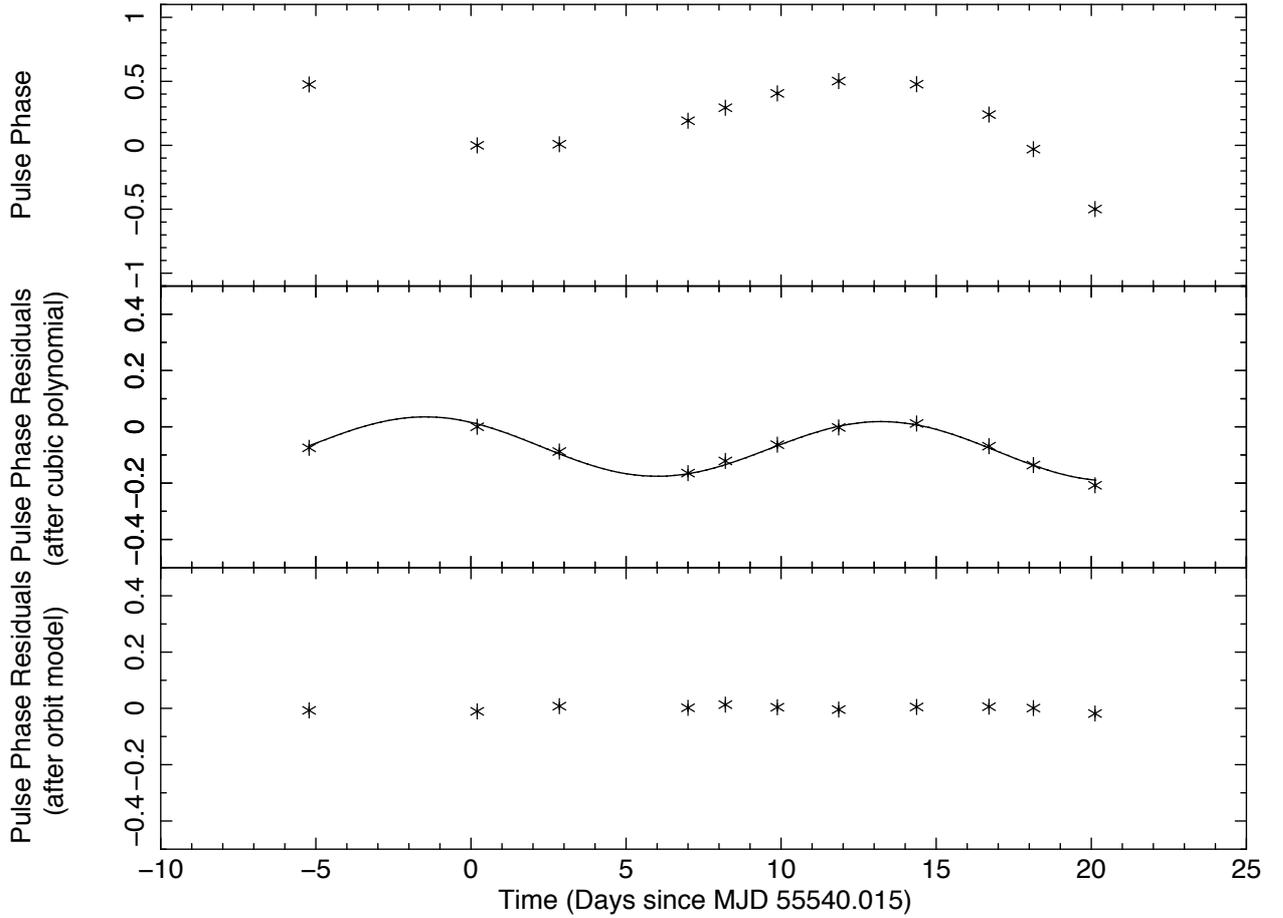}
	\caption{Upper panel: Pulse phases of MAXI J1409$-$619. Middle panel: 
	Residuals of pulse phases after removal of a cubic polynomial. 
	Lower panel: Residuals of pulse phases after removal of the orbit model. Errors are smaller than data marks.} 
	\label{fig:timing}
\end{figure}

The luminosity decreases almost linearly with time during the outburst
as $L\sim L_{0}(-t/t_{0})$. 
As the pulse frequency derivative exhibits a power law dependence on luminosity in accretion disc models (e.g. \cite{GhoshLamb79, Torkelsson1998} and also see Section \ref{sect:torquelum} for more detailed discussion on disc accretion) as $\dot\nu \propto L^{\gamma} \approx L_{0}^{\gamma}(-t/t_{0})^{\gamma}$ where $\gamma$ is the power law index, pulse frequency derivative is affected accordingly, implying a non-zero second time derivative $\ddot\nu$. Therefore a cubic polynomial fitting to the pulse arrival
times describes pulse arrival times better than a quadratic fitting.
In Fig. \ref{fig:timing}, we present the pulse phases and  residuals of the pulse arrival times after the removal of a cubic polynomial.

The residuals after cubic polynomial fit in Fig. \ref{fig:timing} (middle panel) is consistent with a circular orbital model. In general, a model describing a circular orbit can be represented as: 
\begin{equation} 
\delta t_{{\rm{orb}}} = \frac{a_{x} \sin i}{c} \sin\bigg(\frac{2\pi (t-T_{\pi/2})}{P_{orb}} +\frac{\pi}{2}\bigg)
\end{equation}
where $i$ is the inclination angle between the orbital angular momentum vector and the line of sight, $ \frac{a_{x} \sin i}{c}$ is the light travel time for projected semimajor axis, $T_{\pi/2}$ is the orbital epoch where the mean orbital longitude, $ \frac{2\pi (t-T_{\pi/2})}{P_{{\rm{orb}}}} +\frac{\pi}{2}$ , is maximum (i.e. 90$^{\circ}$) and $P_{{\rm{orb}}}$ is the orbital period. The residuals are well fitted to the 
circular orbital model with a period of 14.7(4) days. The orbital model has decreased the
reduced $\chi^2$ from $\sim35$ to $\sim1$.
 The timing and possible orbital parameters obtained by the models are listed in Table \ref{table:timing}. Using these orbital parameters, the mass function 
of the system results as;
\begin{equation}
f(M)=\frac{4\pi ^{2}(a_{x}\sin i)^{3}}{GP^{2}_{orb}}
=\frac{(M_{c} \sin i )^{3}}{(M_{x}+M_{c})^{2}}
\sim 0.1 M_{\odot},
\end{equation}
where $G$ is the gravitational constant, $M_{x}$ is the mass of the neutron star and $M_{c}$ is the mass of the companion star.
A mass function of $\sim0.1 M_{\odot}$ indicates that MAXI J1409$-$619 should have a small inclination angle when we consider the probable O/early B-type companion \citep{Orlandini2012}.  It should be noted that the data used for orbital modelling extends only to $\sim 25$ days (see Fig. \ref{fig:timing}) and therefore, future observations are required to ascertain the suggested orbital model above.

\begin{table}
	\centering
	\caption{Timing solution of MAXI J1409$-$619. A number given in parentheses is the 1$\sigma$ uncertainty in the least significant digit of a stated value.}
	\large{	
			\begin{tabular}{|l|l|}
				\hline \hline
				{\bf{Parameter}} & {\bf{Value}} \\
				\hline 
				Epoch (Days in MJD) & 55540.0150(6) \\
				$\nu$ ($10^{-3}$ Hz) & 1.98287(4) \\
				$\dot{\nu}$ ($10^{-11}$ Hz s$^{-1}$) & 1.599(9) \\
				$\ddot{\nu}$ ($10^{-18}$ Hz s$^{-2}$) & -6.4(8) \\
				Orbital Period (Days) & 14.7(4) \\
				$a/c \sin i$ (s) & 25.7(5) \\
				Orbital Epoch (Days in MJD) & 55547.5375(6) \\
				\hline \hline
	\end{tabular}}
	\label{table:timing}
\end{table}

Then, each two consecutive pulse arrival times are fit with a linear model since the slope of the linear models, $\delta \nu = \delta \phi / (t-t_0)$, enable us to compute the pulse frequencies corresponding to the mid-times of the fitted pulse arrival times.
In Fig. \ref{fig:freq}, we present the pulse frequencies computed via this procedure together with the pulse frequencies measured through \textit{Fermi}/GBM observations by \textit{Fermi}/GBM pulsar project team.

\begin{figure}
	\includegraphics[height=\columnwidth, angle=-90]{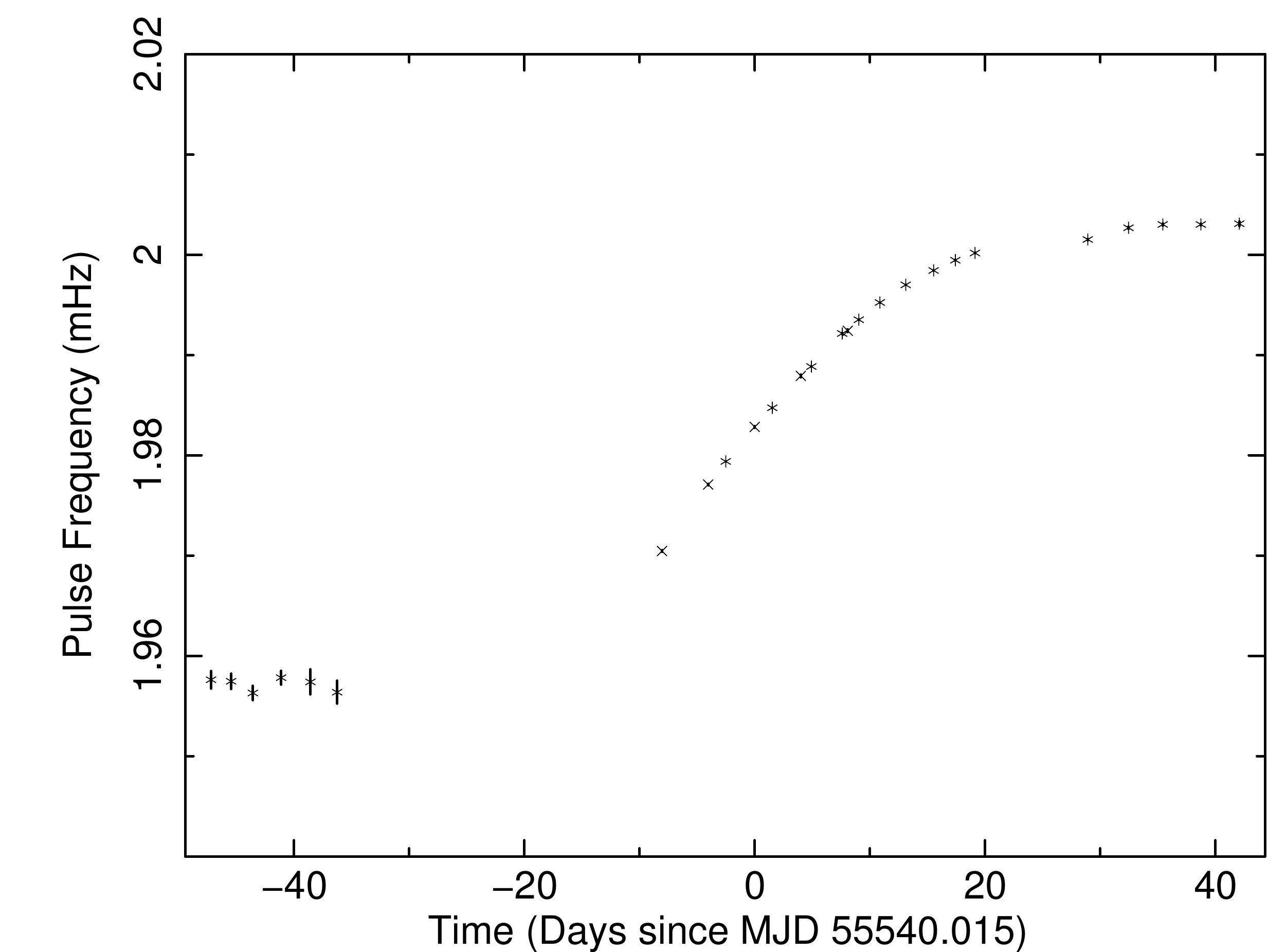}
	\caption{Pulse frequency measurements of MAXI J1409$-$619 deduced from pulse timing analysis of \textit{RXTE}/PCA data (stars) together with the pulse frequencies measured by \textit{Fermi}/GBM pulsar project team (crosses).}
	\label{fig:freq}
\end{figure}

As an alternative to the aforementioned orbital model, the residuals of the cubic polynomial fit can also be interpreted as
 the noise process due to random torque fluctuations
\citep{Bildsten1997, Baykal2007}. We estimate the noise
strength by using the root mean square (rms) residuals of cubic polynomial fit
\citep{Cordes1980, Deeter1984}. In general, for the $r^{th}$-order red noise 
(or the $r^{th}$ order integral of white noise) with strength S$_{r}$, 
the mean square residual for data spanning an interval
with length T is proportional to S$_{r}$T$^{2r-1}$.
The expected mean square residual, after removing a polynomial of degree m
over an interval of length T, is given by
\begin{equation} 
<\sigma _{R}^{2}(m,T)> = S_{r}T^{2r-1}<\sigma _{R}^{2}(m,1)>_{u},
\end{equation}
where the $<\sigma _{R}^{2}(m,1)>_{u}$
is the normalization constant for unit time-scale which can be estimated either by analytical evaluations  or numerical simulations \citep{Deeter1984, Cordes1980, Scott2003}. 
We use m=3 for cubic polynomial and r=2 for second-order red
noise in pulse arrival times (or pulse phases) which corresponds to first
order red noise in pulse frequency.
 We estimate the normalization constant by Monte Carlo simulation of observed 
time series for
 second-order red noise process ($r=2$), for a unit noise strength
S($r=2$). 
Our expected normalization constants are consistent with those obtained 
by direct mathematical evaluation \citep{Deeter1984, Cordes1980}. Using this technique, we found the noise strength associated with torque fluctuations for MAXI J1409$-$619 as $1.3 \times 10^{-18}$ Hz$^2$ s$^{-2}$ Hz$^{-1}$ which is  consistent with other accretion powered pulsars \citep{Baykal1993, Bildsten1997}.

\subsubsection{Torque--luminosity correlations}
\label{sect:torquelum}

Using \textit{RXTE}/PCA observations between MJD 55531 and MJD 55575 corresponding to the 
high luminosity region of the lightcurve, for which a temporary accretion disc formation is plausible, we investigate the relation between 
torque and X-ray luminosity of the source using pulse frequency derivative and 
count rate measurements conducted for each two-day long interval. 

If the source accretes via an accretion disc, the standard accretion disc models  \citep{GhoshLamb79,Ghosh94,Torkelsson1998,Kluzniak2007} suggests a
relation between total external torque exerted on the neutron star and the bolometric
luminosity of the source. For this case, the radius of the inner edge of the accretion disc ($r_0$)
gets smaller as the accretion rate ($\dot{M}$) increases and can be 
approximated as \citep{Ghosh1979b,Pringle72}

\begin{equation}
r_0=K\mu^{4/7}(GM)^{-1/7}{\dot{M}}^{-2/7},
\label{eq:r0}
\end{equation}

{\noindent{where $\mu$ is the magnetic moment of the neutron star ($\mu\simeq	BR^3$, 
$B$ is the surface magnetic field and $R$ is the radius of the neutron star),
$G$ is the universal gravitational constant, $M$ is the mass of the neutron star, and $K$ is a dimensionless parameter of the order of unity.}} 

Total external torque estimate due to accretion can then be expressed as \citep{GhoshLamb79}:
\begin{equation}
2\pi I {\dot{\nu}} = n(\omega_s){\dot{M}}l_K,
\label{eq:torque}
\end{equation} 

{\noindent{where $l_K=(GMr_0)^{1/2}$ is the angular momentum per 
unit mass added by the disc to the neutron star at radius $r_0$ and $I$ 
is the moment of inertia of the neutron star. In this equation, $n(\omega_s)$, 
called "dimensionless torque", is approximately given by \citep{GhoshLamb79, Klus2014}} 

\begin{equation}
n(\omega_s)\approx 1.4(1-\omega_s/\omega_c)/(1-\omega_s).
\label{eq:dimt}
\end{equation}

{\noindent{In the above equation, $\omega_s$ is the fastness parameter, which is equal to
the ratio of the spin frequency of the neutron star to the Keplerian frequency at
$r_0$ radius which can further be expressed as}}

\begin{equation}
\omega_s=\nu/\nu_K(r_0)=2\pi K^{3/2}P^{-1}(GM)^{-5/7}\mu^{6/7}{\dot{M}}^{-3/7},    
\end{equation} 

{\noindent{$P$ is the spin period of the neutron star, $\nu_K(r_0)$ is the Keplerian angular frequency at $r_0$ and $\omega_c$ is the critical fastness parameter with a value of $\sim 0.35$.}}

Due to the release of the gravitational potential energy of the accreted material, 
X-ray emission with a luminosity of

\begin{equation}
L=GM{\dot{M}}/R=\eta \dot{m}c^2, 
\label{eq:lum}   
\end{equation}

{\noindent{occurs at the neutron star surface. In this equation, $\eta$ is a dimensionless efficiency 
factor which is about 0.1 for neutron stars. Using Eqns. \ref{eq:r0}, \ref{eq:torque} and \ref{eq:lum}, the relation between spin frequency derivative
and X-ray luminosity can be written as,}}

\begin{equation}
\dot{\nu}={{n(\omega_s)} \over {2\pi I}}K^{1/2}(GM)^{-3/7}\mu^{2/7}R^{6/7}{L^{6/7}}.
\label{eq:freqderlum}
\end{equation} 

\begin{figure}
	\includegraphics[width=\columnwidth]{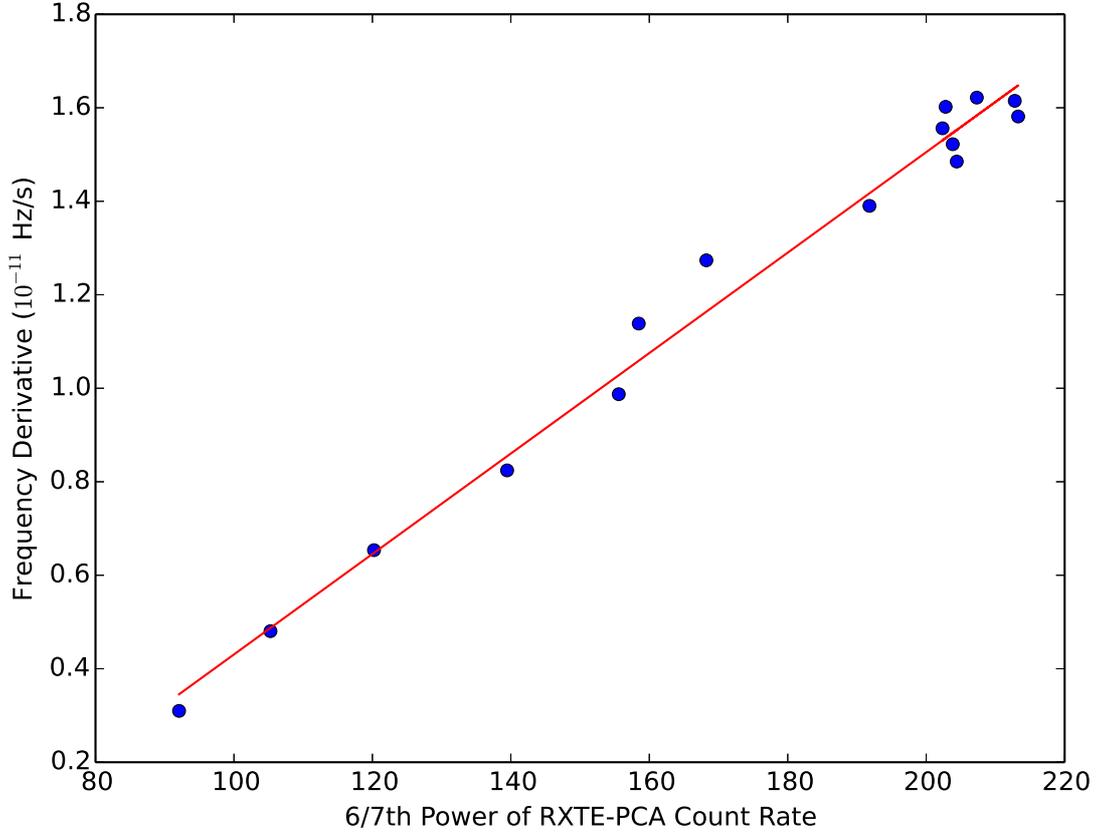}
	\caption{Measured frequency derivatives from \textit{RXTE}/PCA observations as 
	a function of 6/7$^{th}$ power of 2--60 keV \textit{RXTE}/PCA count rate. The solid line 
	indicates the best linear fit.}
	\label{fig:torquelum}
\end{figure}

Eqn. \ref{eq:freqderlum} indicates a linear relationship  between frequency derivative and 6/7$^{th}$ power of luminosity in the presence of strong material torques that give rise to a positive dimensionless torque close to unity. In Fig. \ref{fig:torquelum}, it is evident that frequency derivative is
directly proportional to the 6/7$^{th}$ power of X-ray count rate with a slope of $(1.07\pm0.05)\times 10^{-13}$ (Hz/s)(cts/s)$^{-6/7}$ which is in accord with Eqn. \ref{eq:freqderlum}  with the assumption that the count rate from 2--60 keV band is proportional to the bolometric X-ray luminosity. To relate Eqn. \ref{eq:freqderlum} and Fig. \ref{fig:torquelum}, 2--60 keV count rate of 350 counts/s is found to correspond to a luminosity of about $10^{38}$ ergs/s for a distance of 15 kpc using \textsc{WEBPIMMS}\footnote{\url{https://heasarc.gsfc.nasa.gov/cgi-bin/Tools/w3pimms/w3pimms.pl}} using the absorbed power law model with with the photon index and hydrogen column density of $0.87^{+0.29}_{-0.19}$ and $(2.8^{+3.4}_{-2.2})\times 10^{22}$ cm$^{-2}$ respectively \citep{Orlandini2012}.  

In Eqn. \ref{eq:freqderlum}, the proportionality constant is a function of $n(\omega_s)$ and $B$ for given constant values of the neutron star mass $1.4M_{\odot}$, radius  $10^6$ cm, dimensionless parameters $K$ unity, and moment of inertia $10^{45}$ g cm$^2$. Then, the value of B that minimizes the difference between the calculated slope and the measured slope $(1.07\pm0.05)\times 10^{-13}$ (Hz/s)(cts/s)$^{-6/7}$  is obtained by iteration (i.e. searching numerically for the B value) as $(2.67\pm 0.44)\times 10^{11}$ Gauss yielding a $n(\omega_s)$ value of $1.397\pm 0.001$. From Eqn. \ref{eq:lum}, mass accretion rate ($\dot{M}$) ranges from about $2.61\times 10^{17}$ g s$^{-1}$ to about $7.41\times 10^{17}$ g s$^{-1}$. Using the obtained magnetic field value, lower and upper values of $\dot{M}$ corresponds to inner disc radius estimations ($r_0$) of about $1.21\times 10^8$ cm and $0.90\times 10^8$ cm respectively.

\subsubsection{QPOs} 

To obtain power spectra from the \textit{RXTE}/PCA lightcurve, \textsc{powspec}\footnote{\url{https://heasarc.gsfc.nasa.gov/lheasoft/ftools/fhelp/powspec.txt}} tool 
is used.
We generate the power spectra from short time segments and average them to increase the statistical significance of the spectra, since using long time scales would increase the random scattering of the power estimates \citep{vanderKlis87}.
The lightcurve is divided 
into 256 s long time intervals, each of which contains 1024 bins with 0.25 
second rebinning time.  Each consecutive 9 intervals are averaged into one 
frame; therefore, each power spectrum plot spans 2304 seconds. Then, the results 
are rebinned by a factor of 4. Analogous to other X-ray pulsars, the shape of a single power spectrum of the source can be defined by a broken power law model. 
However, since many power spectra are combined into a single frame, variation in the break frequency with luminosity may smear out and smoothen the sharp transition between two power law components (\citet{Monkkonen2019} and references therein).
Therefore, the continuum of the power spectra is represented by a smoothly broken power law model, which is defined as

\begin{equation}
f(x)=A\bigg(\frac{x}{x_b}\bigg)^{-\alpha_1}\Bigg\{\frac{1}{2}\Bigg[1+\bigg(\frac{x}{x_b}\bigg)^\frac{1}{\Delta}\Bigg]\Bigg\}^{(\alpha_1-\alpha_2)\Delta}
\label{eq:sbpl}
\end{equation}
where $A$ is the amplitude, $x$ is the frequency, $x_b$ is the break 
frequency, $\alpha_1$ and $\alpha_2$ are the power law indices at frequency 
$f<x_b$ and $f>x_b$, respectively, and $\Delta$ is the smoothness parameter 
\footnote{\url{https://docs.astropy.org/en/stable/api/astropy.modeling.powerlaws.SmoothlyBrokenPowerLaw1D.html}}. 
The QPOs are identified as excess peaks in the continuum and are modelled as 
Lorentzian components added to the continuum model. The additive 
Lorentzian component is defined as

\begin{equation}
L=\frac{l_n}{1+\Big(2\times\dfrac{x-l_c}{l_w}\Big)^2}
\label{eq:lor}
\end{equation}
where $x$ is the frequency, $l_n$ is the line normalization parameter, $l_c$ 
is the center frequency of the line, and $l_w$ is the line width (at full-width half-maximum). Appearing 
harmonic peaks of QPO frequency are also modelled with additional Lorentzian components. 
Fitting procedure is carried out using Python libraries\footnote{\url{https://docs.scipy.org/doc/scipy/reference/generated/scipy.optimize.curve_fit.html}}, and we obtain the best parameters and their error ranges at 1$\sigma$ confidence level.

The significance of the QPO features are 
calculated by normalizing the power spectrum via dividing it by the modelled continuum 
and multiplying the result by 2 \citep{vanderKlis1988}. With the continuum normalized to 2, a rebinning factor of 4 and the number of averaged intervals of 9, the resulting power 
spectrum continuum would be consistent with a $\chi^2$ distribution for 
$2\times4\times9=72$ degrees of freedom (dof). As an example, a signal at an excess power 
of 4 (see Figure \ref{fig:qpo2-07-01}, bottom panel) corresponds to a total power of $4\times4\times9=144$, and the 
probability of having a false signal, $P$, is calculated as 
$P(144|72)=1.01\times10^{-6}$. We have 252 frequencies in each power 
spectra; therefore, the total probability of having a false signal is calculated 
as $252\times1.01\times10^{-6}=2.53\times10^{-4}$, which corresponds to 
$3.66\sigma$ level of detection (see \citet{Inam2004} and \citet{Acuner2014} for examples). We select the QPOs and their harmonics which have
more than $3\sigma$ significance, seen in Figure~\ref{fig:qpofreqflux}. One 
example of a power spectrum fit and corresponding QPO and its harmonic are shown in 
Figure~\ref{fig:qpo2-07-01}. Furthermore, in order to indicate the coherence of oscillation signals (for QPOs), we also  
calculate the quality factors (Q-value) of corresponding Lorentzian components, simply defined as the ratio of the line width to the centroid frequency, which are also given in Table \ref{table:qpoparam}. The calculated quality factors are well above the broadband noise level ($Q>2$, see e.g. \cite{Acuner2014}) and indicate clear detection of QPO signals.

\begin{figure}
	\includegraphics[width=\columnwidth]{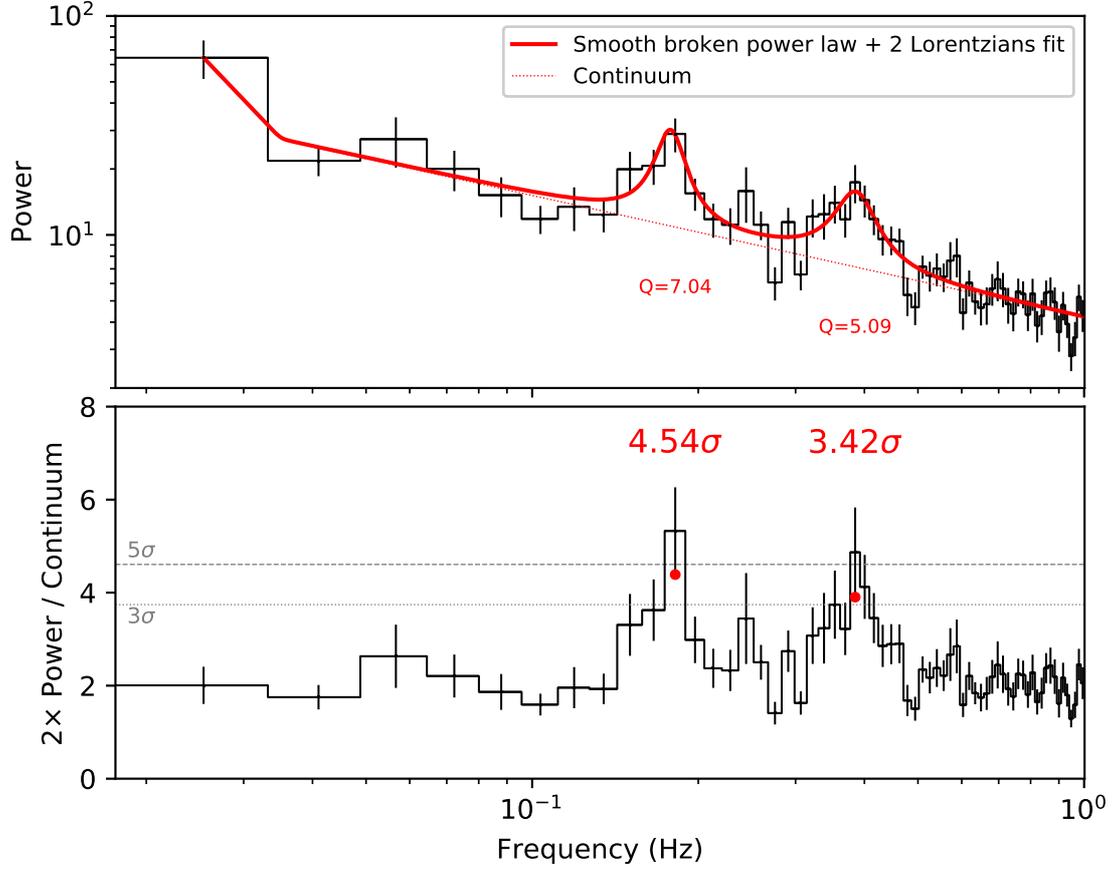}
	\caption{Upper panel: The power spectrum of the first 2304-second-long segment of 
	observation 95441-01-01-06 (MJD 55545). The dashed red line indicates the fit for the continuum and the solid red line represents the fit after the addition of Lorentzian components. Q-values of the Lorentzian components are shown as well. Lower panel: The power spectra after normalization to the continuum and the detection levels of the excess signals (see the text below). The dashed lines indicate 3$\sigma$ and 5$\sigma$ detection levels for the possible QPO structures.}
	\label{fig:qpo2-07-01}
\end{figure}

\begin{figure}
	\includegraphics[width=\columnwidth]{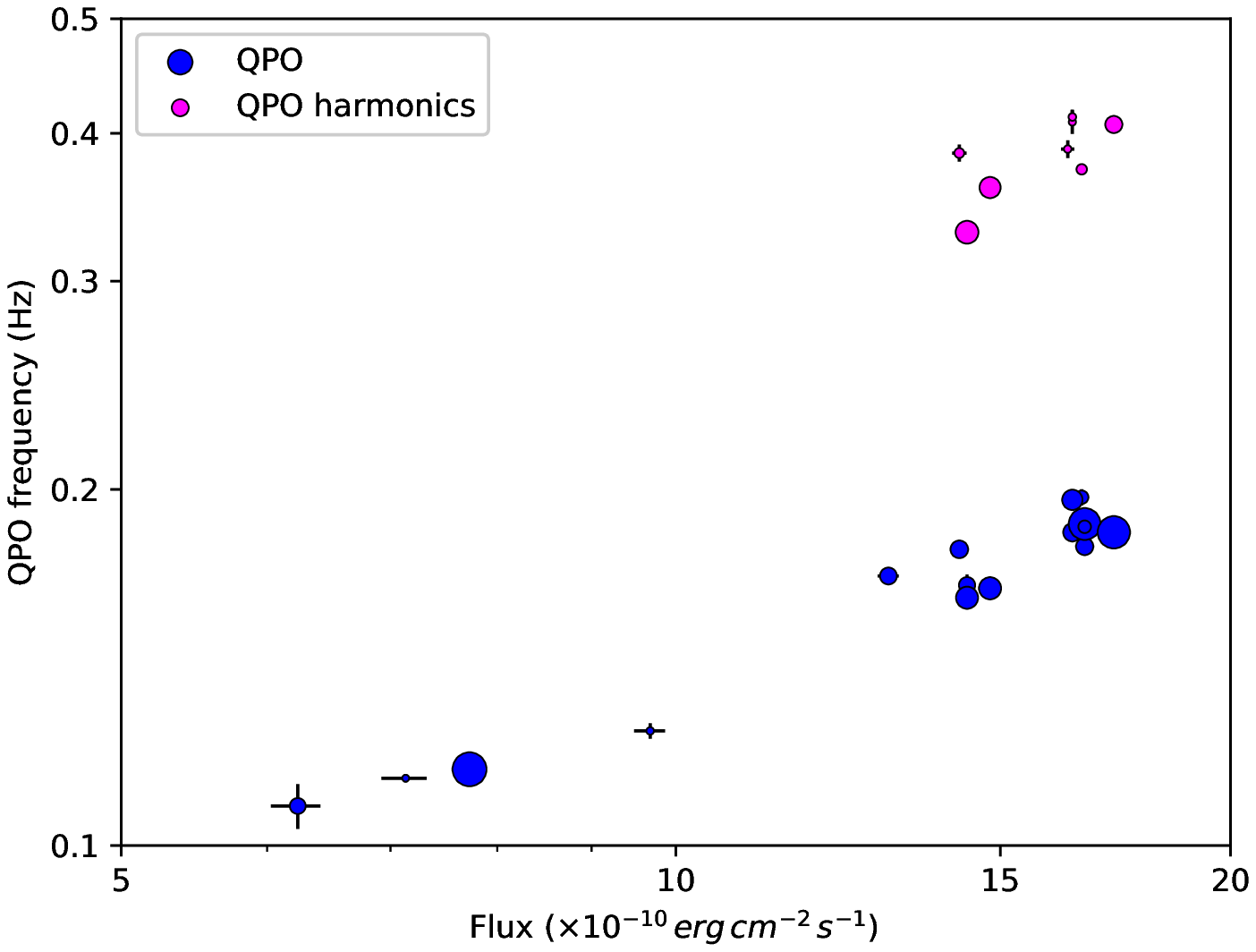}
	\caption{QPO centroid frequencies (blue circles) and QPO harmonics (magenta circles) versus flux. QPOs with only more 
	than $3\sigma$ significance are included. The magnitude of the circular marks implies the significance level of the QPOs, and the biggest circle in the plot 
	corresponds to $6.90\sigma$. The errors smaller than the corresponding circles are not visible in the figure (see Table \ref{table:qpoparam} for the QPO centroid frequencies).}
	\label{fig:qpofreqflux}
\end{figure}

QPOs are detected only at the first half of the decaying part of the outburst in 2010 December, and the signal possesses a transient nature, occasionally disappearing altogether in the power spectra.
QPO harmonics seem to appear for the most cases in which the principal Lorentzian component is observed, found around double the centroid frequency of the main QPO features, 
though many of these harmonics are not statistically significant. 

Another notable result is that the QPO centroid frequencies decrease consistently over time. The first QPOs are 
detected around 0.2 Hz; however, as the outburst flux decays, QPO centroid frequencies decrease down to 0.1 Hz (see Figure~\ref{fig:qpofreqflux}). As the source flux consistently decreases 
over the time period in which the QPOs are detected, we can conclude that QPO centroid frequencies are correlated with flux, similar to the case of A0535+262 \citep{Finger1996x}.

\begin{table}
	\centering
	\caption{Detected QPOs of MAXI J1409$-$619.}
	\begin{tabular}{cl|cc|rr|cc}
		\multirow{2}{*}{Obs. ID} & \multicolumn{1}{c|}{Epoch} & \multicolumn{2}{c|}{$\nu_{QPO}$ (Hz)}             & \multicolumn{2}{c|}{Q-value}  				   & \multicolumn{2}{c}{Significance ($\sigma$)}	   \\
		 & \multicolumn{1}{c|}{(MJD)} & \multicolumn{1}{c}{\#1} & \multicolumn{1}{c}{\#2} & \multicolumn{1}{c}{\#1} & \multicolumn{1}{c}{\#2} & \multicolumn{1}{c}{\#1} & \multicolumn{1}{c}{\#2} \\ \hline
\multicolumn{1}{c}{\multirow{2}{*}{95441-01-01-02}} & 55541.621                  & 0.197$\pm$0.003 & -               & 6.340                   & -                        & 3.934               & -                   \\
\multicolumn{1}{c}{}                                & 55541.648                  & 0.189$\pm$0.002 & 0.373$\pm$0.003 & 6.751                   & 18.672                   & 3.103               & 3.510               \\ \hline
\multirow{3}{*}{95441-01-01-00}                     & 55541.804                  & 0.196$\pm$0.003 & 0.409$\pm$0.009 & 7.788                   & 4.311                    & 4.934               & 3.044               \\
& 55541.831                  & -               & 0.413$\pm$0.006 & -                       & 20.630                   & -                   & 3.103               \\
& 55541.957                  & 0.184$\pm$0.002 & -               & 8.924                   & -                        & 4.636               & -                   \\ \hline
\multirow{3}{*}{95441-01-01-03}                     & 55542.665                  & 0.179$\pm$0.003 & -               & 3.620                   & -                        & 4.491               & -                   \\
& 55542.709                  & 0.187$\pm$0.001 & -               & 8.862                   & -                        & 6.590               & -                   \\
& 55542.785                  & 0.186$\pm$0.002 & -               & 4.279                   & -                        & 3.822               & -                   \\ \hline
\multirow{2}{*}{95441-01-01-04}                                   & 55543.818                  & 0.184$\pm$0.001 & -               & 7.895                   & -                        & 6.667               & -                   \\
& 55543.845                  & -               & 0.407$\pm$0.003 & -                       & 19.310                   & -                   & 4.467               \\ \hline
95441-01-01-05                                      & 55544.798                  & -               & 0.388$\pm$0.007 & -                       & 4.580                    & -                   & 3.087               \\ \hline
95441-01-01-06                                      & 55545.638                  & 0.178$\pm$0.002 & 0.385$\pm$0.006 & 7.036                   & 5.089                    & 4.543               & 3.421               \\ \hline
95441-01-01-07                                      & 55546.835                  & 0.165$\pm$0.003 & 0.360$\pm$0.002 & 3.042                   & 11.423                   & 5.191               & 5.045               \\ \hline
\multirow{2}{*}{95441-01-02-00}                     & 55547.097                  & 0.166$\pm$0.003 & 0.330$\pm$0.005 & 2.430                   & 4.644                    & 4.348               & 5.291               \\
& 55547.124                  & 0.162$\pm$0.002 & -               & 6.661                   & -                        & 5.186               & -                   \\ \hline
95441-01-02-03                                      & 55549.395                  & 0.169$\pm$0.002 & -               & 5.096                   & -                        & 4.439               & -                   \\ \hline
95441-01-03-00                                      & 55554.353                  & 0.125$\pm$0.002 & -               & 5.388                   & -                        & 3.089               & -                   \\ \hline
95441-01-03-02                                      & 55556.705                  & 0.116$\pm$0.001 & -               & 4.930                   & -                        & 6.897               & -                   \\ \hline
95441-01-03-03                                      & 55558.142                  & 0.114$\pm$0.001 & -               & 5.253                   & -                        & 3.003               & -                   \\ \hline
95441-01-03-04                                      & 55559.644                  & 0.108$\pm$0.005 & -               & 5.036                   & -                        & 4.294               & -                  
	\end{tabular}
	\label{table:qpoparam}
\end{table}

\subsubsection{Pulsed fractions}

Using \textit{RXTE}/PCA lightcurve, we calculate the pulsed fraction (P.F.) of the source for each 
observation. For our 
calculations, we use 20 binned pulse profiles and use the definition 
$P.F. = (I_{max}-I_{min})/(I_{max}+I_{min})$, where $I_{min}$ and $I_{max}$ are 
the count rates of the bins with the minimum and the maximum count 
rate respectively. To demonstrate the flux dependence of pulsed fraction of the source, 
we plot these measurements as a function of \textit{RXTE}/PCA count rate of the source 
in Fig. \ref{fig:pulsedfrac}. The pulsed fraction exhibits a positive correlation with the count rate up to $\sim$200 cts/s (corresponding to the luminosity of $\sim6\times10^{37}$ erg/s for the source distance suggested by \cite{Orlandini2012}), beyond which the correlation switches to negative.

\begin{figure}
	\includegraphics[height=\columnwidth, angle=-90]{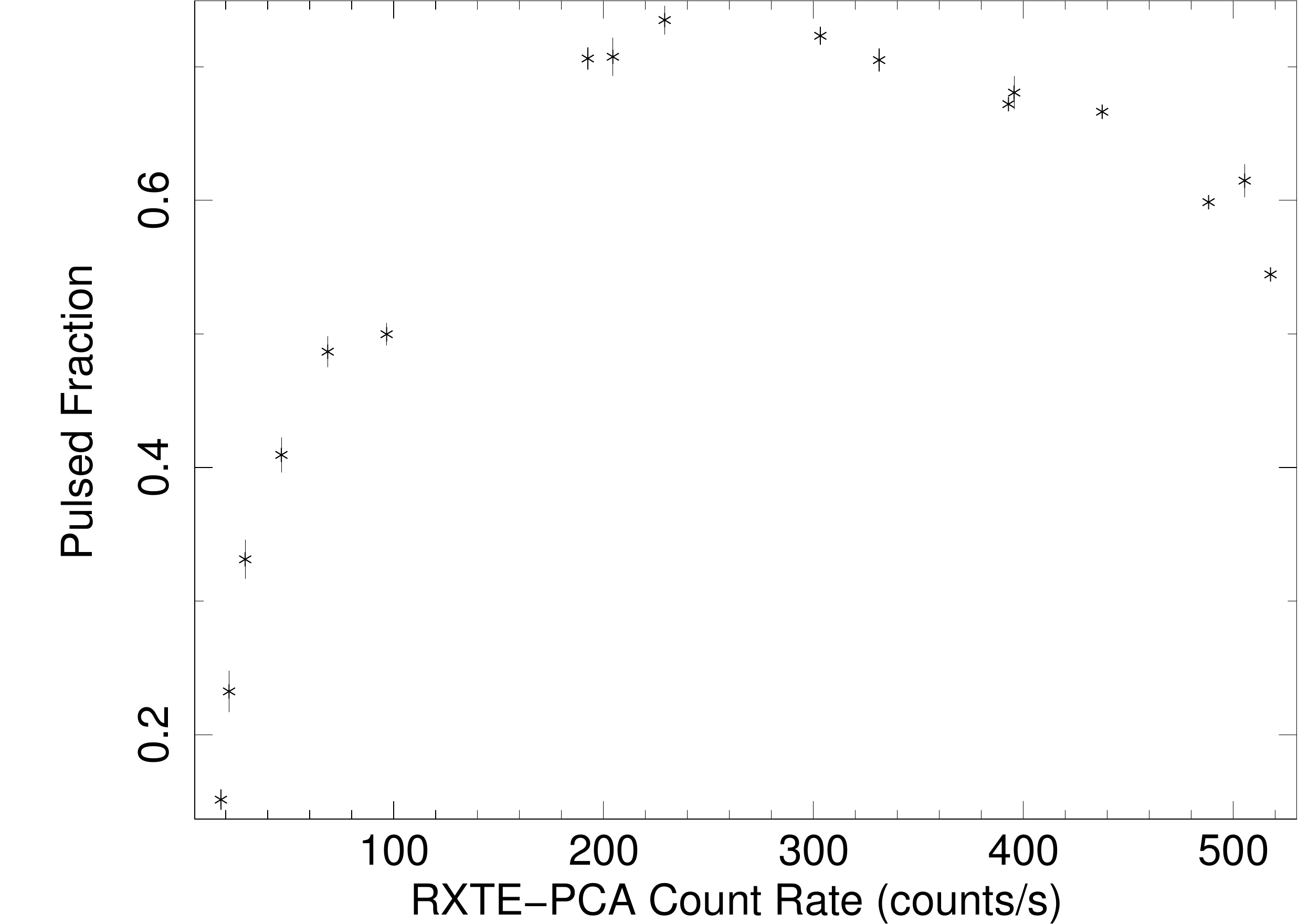}
	\caption{Pulsed fraction as a function of 2--60 keV \textit{RXTE}/PCA count rate of 
	the source.}
	\label{fig:pulsedfrac}
\end{figure}

\subsection{Spectral Analysis}
\subsubsection{Time-resolved spectra}
\label{subsubsec:timeresolvedspectra}
For \textit{RXTE}/PCA spectral analysis, we first combine and fit all observations 
in 95441 set (between MJD 55491--55597 with a total exposure of 74.4 ks) resulting in a single spectrum. While fitting, 3--25 keV band is used, a systematic error of 0.5\% is set, and chi-statistics is utilised. An absorbed power law model with a high energy cutoff plus a Gaussian iron line fixed at 6.4 keV (\textsc{phabs*(cutoffpl+gauss)}) is employed to describe the resulting spectrum (see Figure~\ref{fig:rxtespec95441}). The error ranges of the model parameters are calculated at the 90\% confidence level. For the \textsc{phabs} component, the default solar abundances and photoionization cross-sections predefined in \textsc{xspec} are used (see \citet{Anders1989} and \citet{Verner1996} for the values).

Afterwards, the observations are fitted separately in order to investigate 
the evolution of the fit parameters over time. Subsequent short observations 
are combined before fitting when the exposure times are insufficient, leading to 37 statistically significant spectra. Following the same steps as in the analysis of the combined spectrum of the 95441 set, the spectra are generated within 3-25 keV energy band. While modeling, error range of parameters are calculated at 90\% confidence level and the represented flux values are calculated in 3-25 keV energy band. Most of the spectra are fitted with a model including the high energy cutoff. However, as the count rates of the observations decrease, decreasing signal-to-noise ratio deteriorates the spectral quality, which prevents the fit to converge, and therefore constrains the cutoff energy values. Therefore some spectra are fitted with a power law model without the cutoff model. Additionally, a 0.5\% systematic error is set while fitting spectra of combined or long-duration observations. The evolution of the spectral parameters with time is presented in Figure~\ref{fig:rxtespectimeparam}.

\begin{figure}
	\includegraphics[width=\columnwidth]{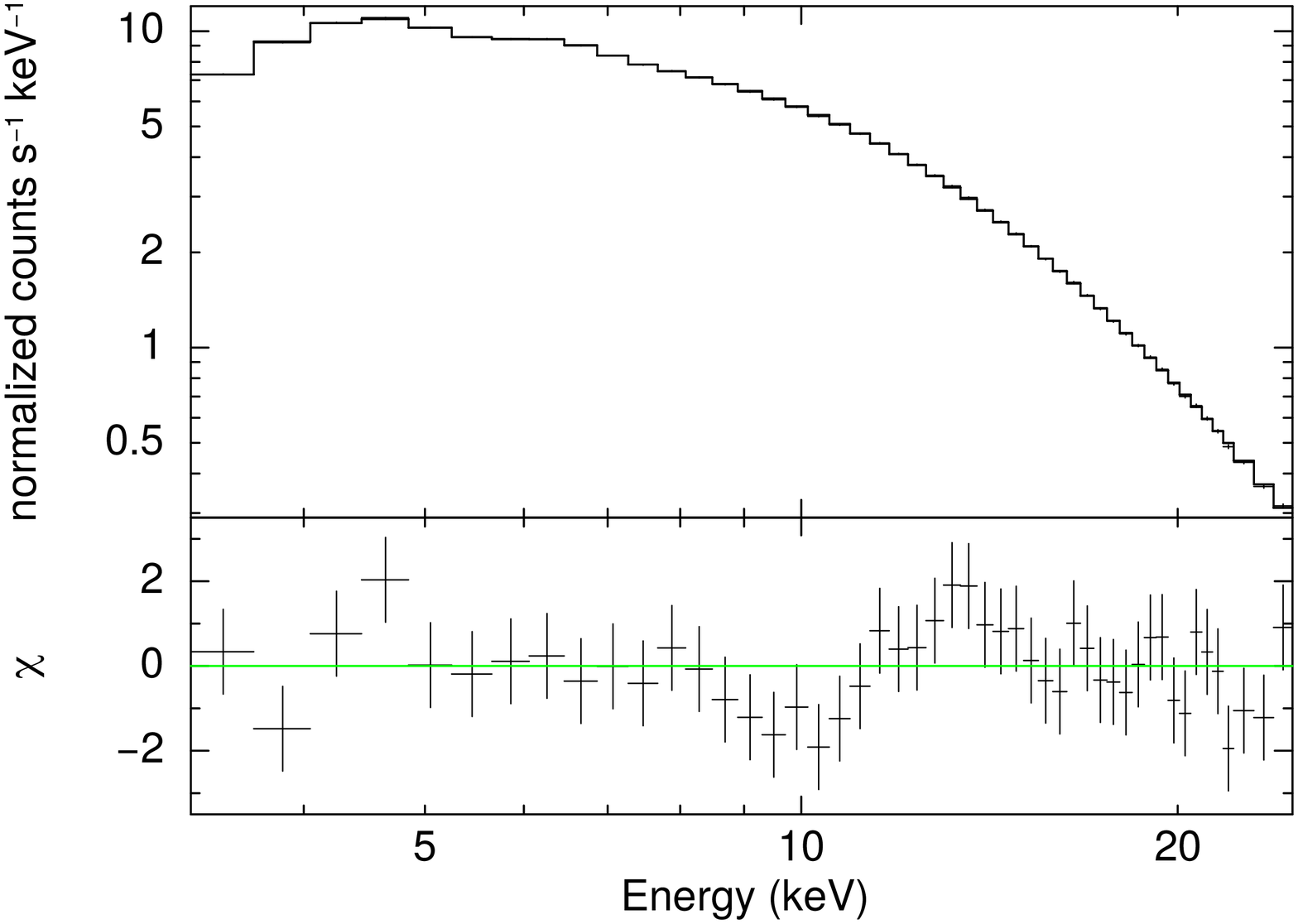}
	\caption{Upper panel: \textit{RXTE}/PCA spectrum of the whole 95441 set (MJD 55540--55560, 74.4 ks) with a power law fit with a high energy cutoff plus a Gaussian iron line fixed at 6.4 keV. Reduced $\chi^2$ of the fit is 1.014. The errors in the spectrum are too small to be visible. Lower panel: Residuals of the fit in $\sigma$ values.}
	\label{fig:rxtespec95441}
\end{figure}

\begin{figure}
	\includegraphics[width=\columnwidth]{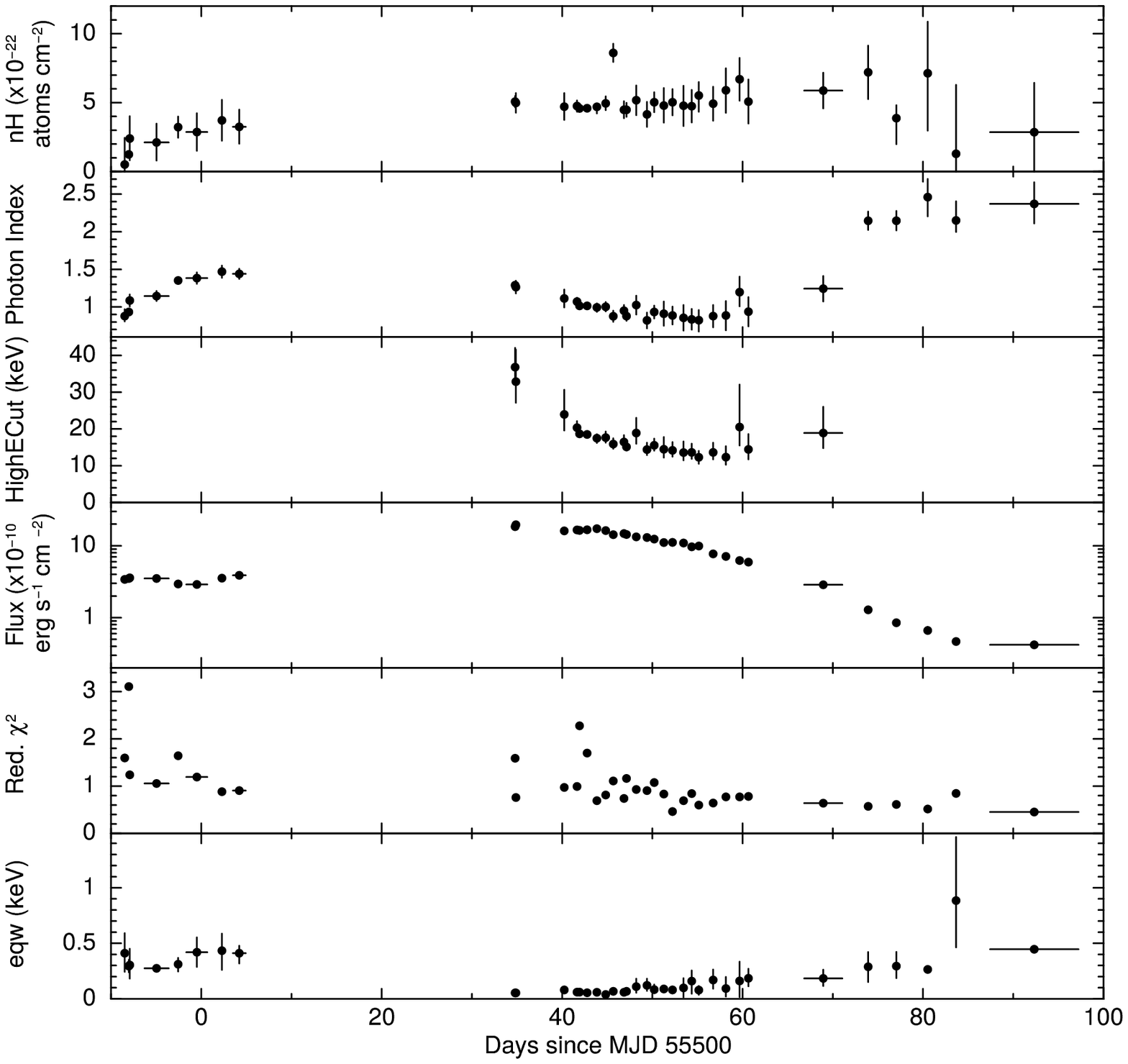}
	\caption{From top to bottom, the panels show the time evolution of the hydrogen column density ($n_H$), photon index, cutoff energy, flux at 3--25 keV energy band, reduced $\chi^2$ and equivalent width of the Gaussian component of \textit{RXTE}/PCA observations. The error bars in flux are not seen as they are smaller than the data marks.}
	\label{fig:rxtespectimeparam}
\end{figure}

In general, the models accurately fit the spectra: Only two spectra have reduced $\chi^2$ 
values exceeding 2. Photon index values exhibit an inverse proportional 
relation with flux up to a certain degree, after which photon index values 
stay relatively flat. This relation is shown in 
Figure~\ref{fig:rxtespectimePIflux}. High energy cutoff values are only 
obtained around the peak of the outburst and are decreasing over time. When MAXI J1409-619 was discovered in 2010 October, and at the tail of the outburst, the cutoff energies are not constrained and the spectra can be equivalently described without the cutoff component. In order to show the unconstrained nature of the cutoff component before the outburst peak, two contour plots before (MJD 55492) and during the outburst peak (MJD 55453) are plotted as in Figure \ref{fig:contour}.
Therefore it should be noted that within these time intervals, the photon indices are obtained without the cutoff model since the cutoff values are generally unconstrained at low flux observations.

\begin{figure}
	\includegraphics[width=\columnwidth]{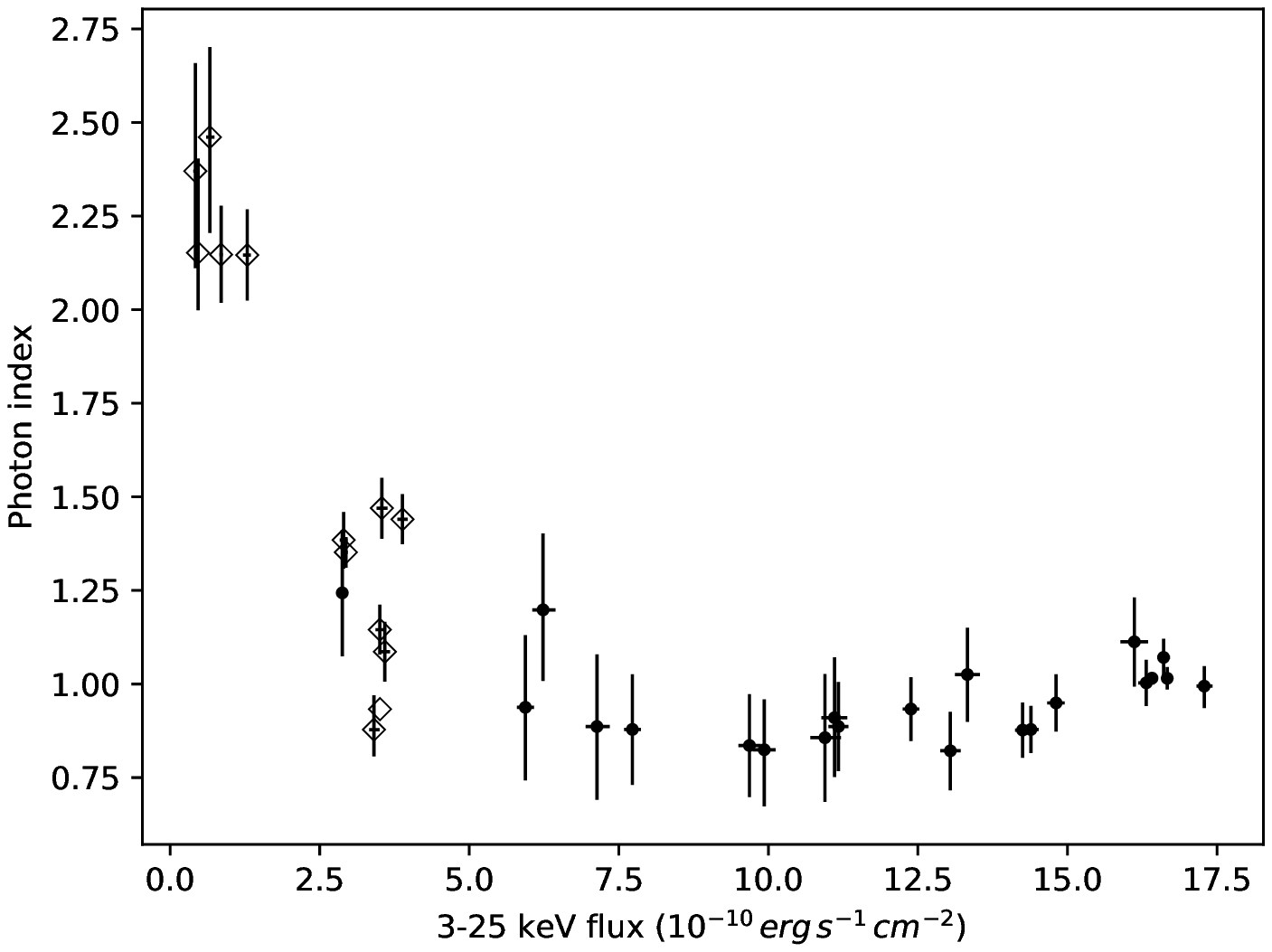}
	\caption{The relation between photon indices and fluxes of \textit{RXTE}/PCA observations. Photon indices obtained from the spectral model without a high energy cutoff component are plotted as diamonds. The critical luminosity of the source resides within a range of $(5-10)\times 10^{-10}$ erg s$^{-1}$ cm$^{-2}$  (see Section \ref{sec:discuss} for details).}
	\label{fig:rxtespectimePIflux}
\end{figure}

\begin{figure}
	\centering
	\includegraphics[width=0.45\columnwidth]{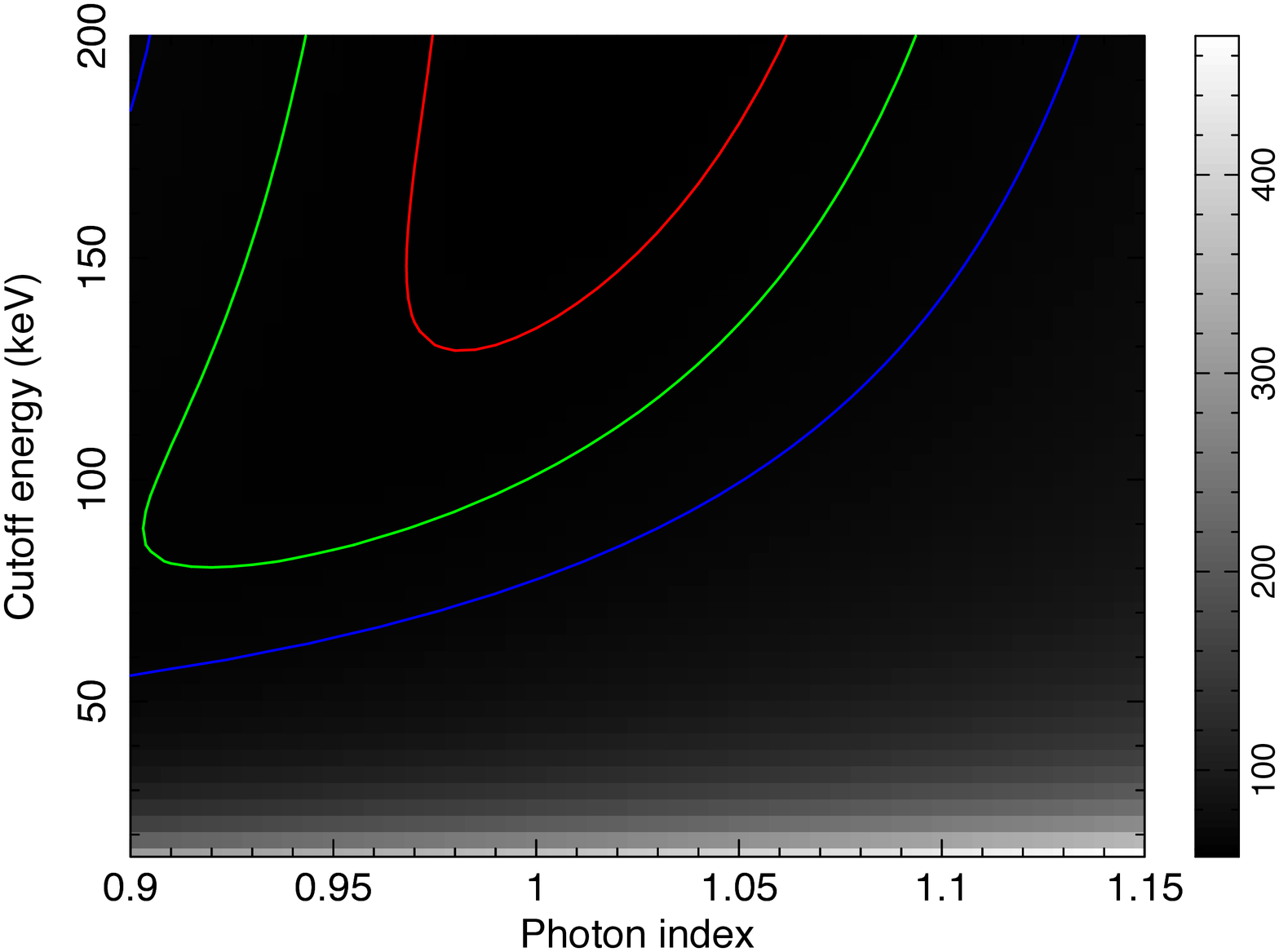}
	\hspace{5mm}
	\includegraphics[width=0.45\columnwidth]{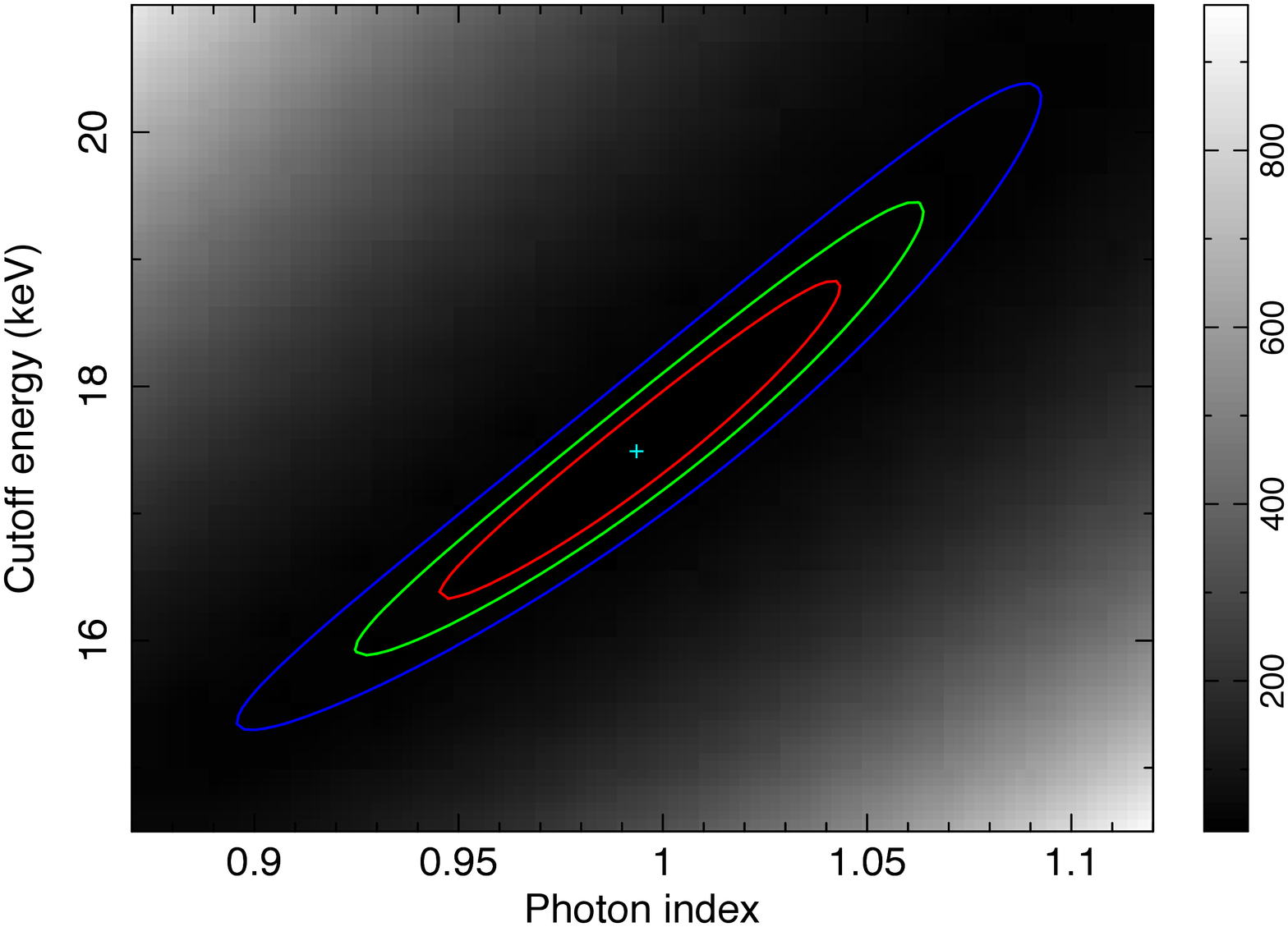}
	\caption{Confidence contour plots in the photon index -- cutoff energy plane, created for RXTE/PCA spectral modelling. The contours correspond to 1$\sigma$, 2$\sigma$, and 3$\sigma$ confidence levels. Sidebar legends show the range of the $\chi^2$ statistics values. Left panel: The cutoff energy parameter is unbounded in the spectral modelling of the observation 95358-02-01-01 (MJD 55492). Right panel: The cutoff energy parameter is well inside the 3--25 keV range in the spectral modelling of the observation 95441-01-01-04 (MJD 55543).}
	\label{fig:contour}
\end{figure}

Especially in long observations, a bump around 10 keV is evident in the residuals of the fits. This feature was mentioned by \citet{Coburn2002} and it was also observed 
in many accreting X-ray pulsars (e.g. Her X-1 and 4U 1907+09), not only in \textit{RXTE} data but also \textit{in Ginga} and \textit{BeppoSAX} data. 
Being always at around 10 keV, the feature is not considered to be a cyclotron line; therefore, it was concluded that it might be an 
intrinsic feature common in accreting pulsars. Even though, its origin is not yet clear, this feature is attributed to the broadening of cyclotron line due to comptonization along column \citep{Farinelli2016}.

\textit{Swift}/XRT has a moderate spectral resolution, and no single 
observation of it has an exposure of more than 2 ks; therefore, observations 
close in time and having similar count rates are combined into 6 groups in order 
to increase the statistical significance of the resulting spectra. Additionally, considering the low count rates in the \textit{Swift}/XRT observations, we proceed with the spectral analysis using C statistics \citep{Cash1979} via
\textsc{cstat} tool within \textsc{xspec}. The spectra are fitted simply with an absorbed 
power law (\textsc{phabs*powerlaw}) model. The error of the model parameters are calculated at the 90\% 
confidence level, and the 0.5--10 keV flux of the entire model is derived via the \textsc{cflux} component. The given spectral model successfully fits the spectra, yet the spectral 
quality is inferior compared to \textit{RXTE}/PCA data (see Figure \ref{fig:swiftspec4} (\textit{Swift}/XRT) and \ref{fig:rxtespec95441} (\textit{RXTE}/PCA) as an example for comparison). 
The evolution of the parameters is consistent with those obtained using the 
analysis of \textit{RXTE}/PCA data. The reduced 
$\chi^2$ values of the fits are between 0.72 and 1.21. Results of spectral fitting of \textit{Swift}/XRT observations are given in Table \ref{table:swift}.

\begin{figure}
	\includegraphics[width=\columnwidth]{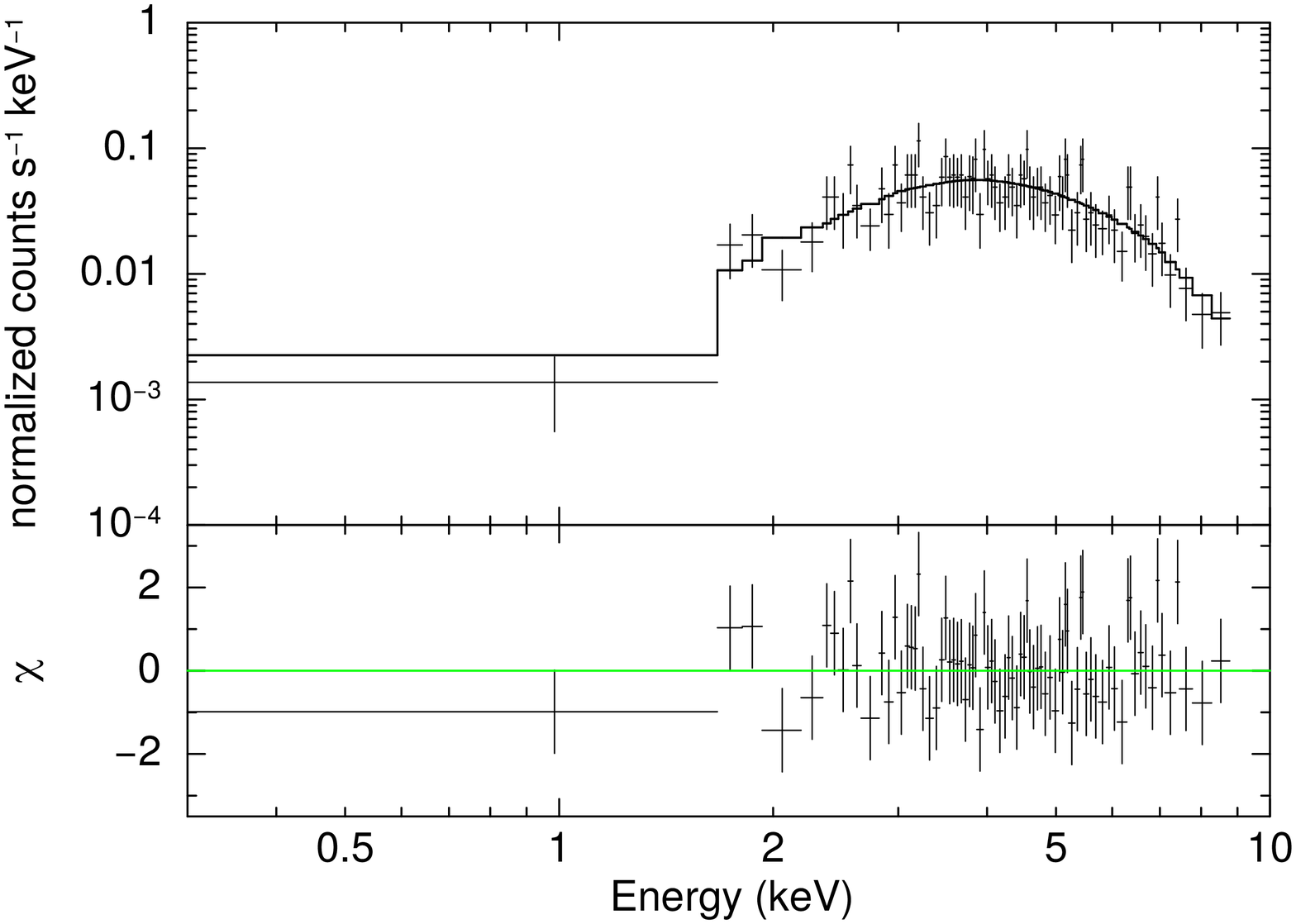}
	\caption{Upper panel: \textit{Swift}/XRT spectrum of MAXI J1409$-$619 with a power law model fit (MJD 55501--55503; Group 4 in Table \ref{table:swift}). Lower panel: Residuals of the fit in $\sigma$ values.}
	\label{fig:swiftspec4}
\end{figure}

\begin{table}
	\centering
	\caption{Spectral parameters of \textit{Swift}/XRT data. The flux values are calculated in 0.5--10 keV energy band. The parameters with asterisk mark (*) are kept fixed during spectral fitting of the given group.}
	\small{
		\begin{tabular}{l||l|l|l|l|l|l|}
			& Group 1                & Group 2                & Group 3                & Group 4                & Group 5                & Group 6                \\ \hline
			\vspace{1mm}                                  
			Obs. midtime \scriptsize{(MJD)}    & $55490.89$             & $55493.19$             & $55498.34$             & $55502.52$             & $55530.86$             & $55583.68$             \\
			\vspace{1mm}
			Exposure (s)                 & $3590$                 & $3305$                 & $2111$                 & $2051$                 & $2597$                 & $4615$                 \\
			\vspace{1mm}
			$n_H$ ($\times10^{22}$ cm$^{-2}$)  & $5.43^{+0.74}_{-0.66}$ & $3.78^{+0.98}_{-0.86}$ & $2.99^{+1.11}_{-0.96}$ & $5.35^{+1.65}_{-1.41}$ & $5.21^{+0.63}_{-0.59}$ & $9.08^{+4.44}_{-3.47}$ \\
			\vspace{1mm}
			Photon index                       & $0.73$*                 & $0.73^{+0.32}_{-0.31}$ & $0.30^{+0.36}_{-0.35}$ & $1.03^{+0.46}_{-0.43}$ & $0.89^{+0.18}_{-0.18}$ & $1.78^{+1.00}_{-0.90}$ \\
			\vspace{1mm}
			Flux  (10$^{-10}$ erg s$^{-1}$ cm$^{-2}$) & $1.54^{+0.13}_{-0.12}$ & $1.31^{+0.13}_{-0.11}$ & $2.15^{+0.22}_{-0.20}$ & $2.01^{+0.47}_{-0.26}$ & $9.98^{+0.67}_{-0.55}$ & $0.21^{+0.49}_{-0.08}$ \\
			\vspace{1mm}
			Reduced $\chi^2$                   & $1.0615$               & $1.1150$               & $0.9286$               & $0.7237$               & $0.99558$              & $1.163$                            
		\end{tabular}}
	\label{table:swift}
\end{table}

\subsubsection{Pulse-phase-resolved spectra}
\label{subsubsection:pulsephasespec}
We perform pulse-phase-resolved spectral analysis for the 3 observations conducted by \textit{RXTE}/PCA} on 2010 December 11--12 (MJD 55541--55542) during the peak of the outburst, totalling a 34.0 ks exposure. A spin period of 503.62 s 
(1.9856 mHz) is used, which is found from timing analysis (see Sec. \ref{sec:timing}). A total of 10 phase bins are 
used; as a result, the spectrum of each phase bin is obtained from about 3.4 ks of exposure. The phase-resolved 
spectra are fitted by an absorbed power law with a high energy cutoff plus a Gaussian iron line fixed at 6.4 keV. Reduced $\chi^2$ values of the phase-resolved spectra range between 0.941 and 1.895.
\begin{figure}
	\centering
	\includegraphics[width=0.48\columnwidth]{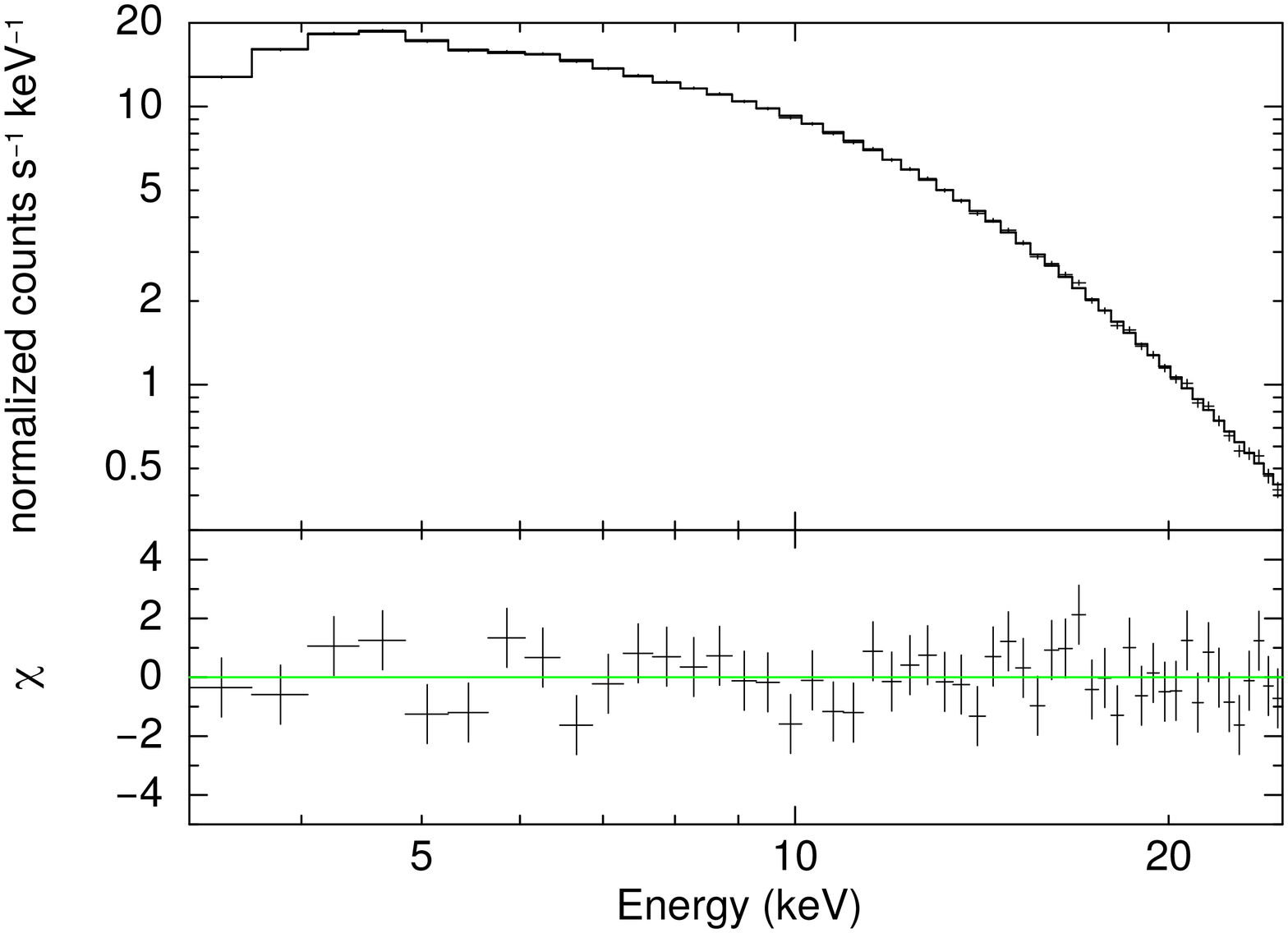}
	\includegraphics[width=0.48\columnwidth]{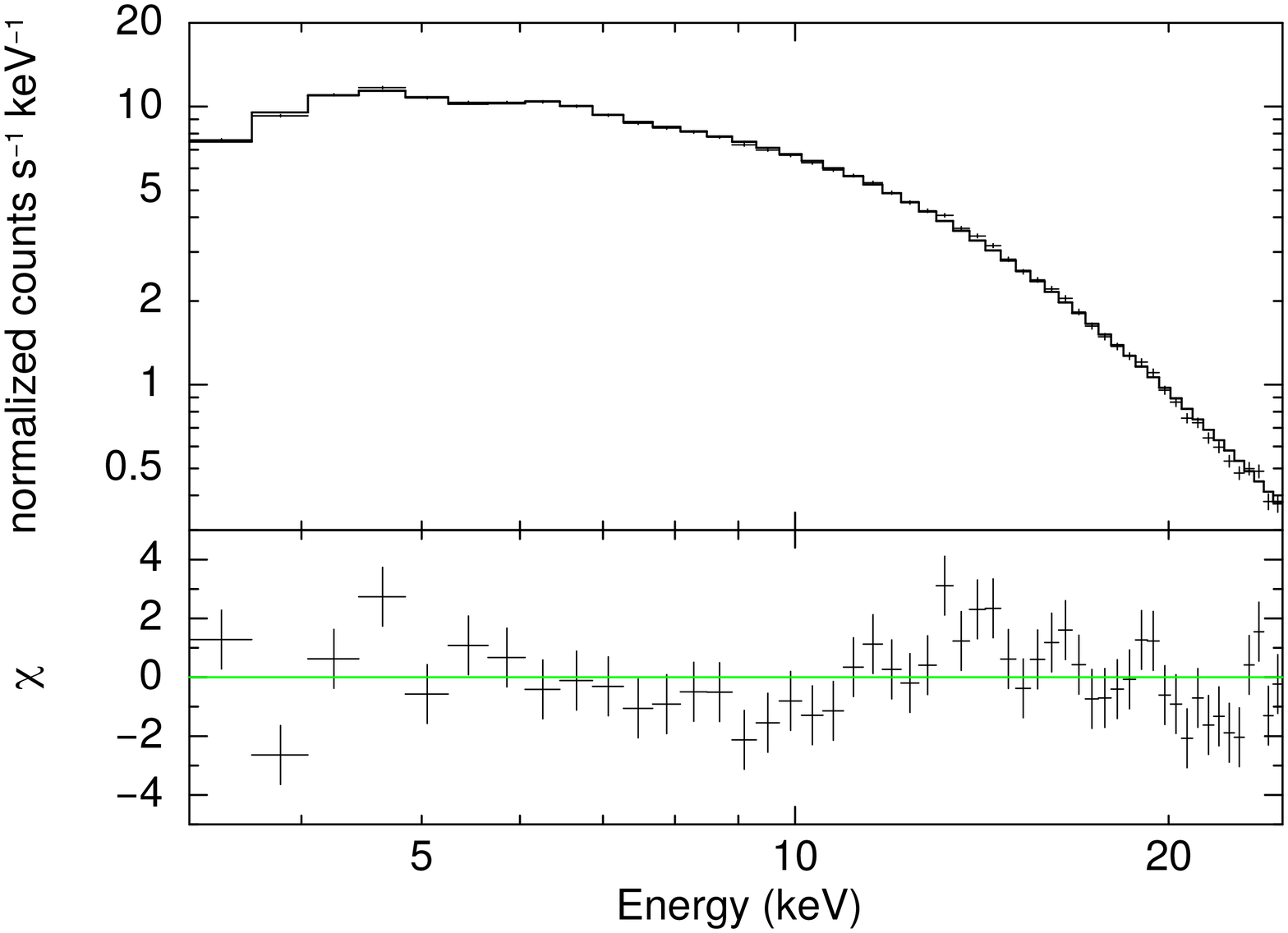}
	\caption{Two fitted phase-resolved \textit{RXTE}/PCA spectra with residuals for comparison. Left panel: The spectrum of phase 0.3 with a low $\chi^2=0.94$. Right panel: The spectrum of phase 0.3 with a high $\chi^2=1.90$.}
	\label{fig:fasebinspecbestworst}
\end{figure}
The pulse-phase dependence of the spectral parameters is presented in  Figure~\ref{fig:fasebinparam}. The double-peaked nature of the pulse profile is  evident 
in phase-dependent flux measurements. The neutral hydrogen column density $n_H$ does not have a 
significant variation with changing phase. The photon index appears to be 
low during the pulse peaks, and the emission softens notably when the flux is minimum. 

\subsubsection{Energy Dependent Pulse Profiles}

In addition to the pulse profile shown in the uppermost panel of Figure \ref{fig:fasebinparam}, the energy dependence of pulse profiles is investigated as well. For this analysis, we use the same 3 RXTE/PCA observations that are also used for the pulse-phase-resolved spectral analysis (see Sec. \ref{subsubsection:pulsephasespec}). These observations, conducted on MJD 55541--55542, have flux levels around $1.6 \times 10^{-9}$ erg cm$^{-2}$ s$^{-1}$. The observations are combined, and the resulting data is folded using the previously found spin period of 503.62 s (see Sec. \ref{sec:timing}). Then, normalized phase profiles with 10 phase bins are created separately for 3--8, 8--13, 13--18, 18--25, and 25--60 keV energy ranges using the \textsc{fbssum} tool.

As it can be seen in Figure~\ref{fig:fasebinenergy}, the pulse profile of MAXI J1409$-$619 do not exhibit strong energy dependence; however, the pulsed fraction alters in different bands. While it is most prominent in 8--13 keV energy band, it diminishes at higher energies, almost vanishing above 25 keV.

\begin{figure}
	\centering
	\includegraphics[width=0.5\columnwidth]{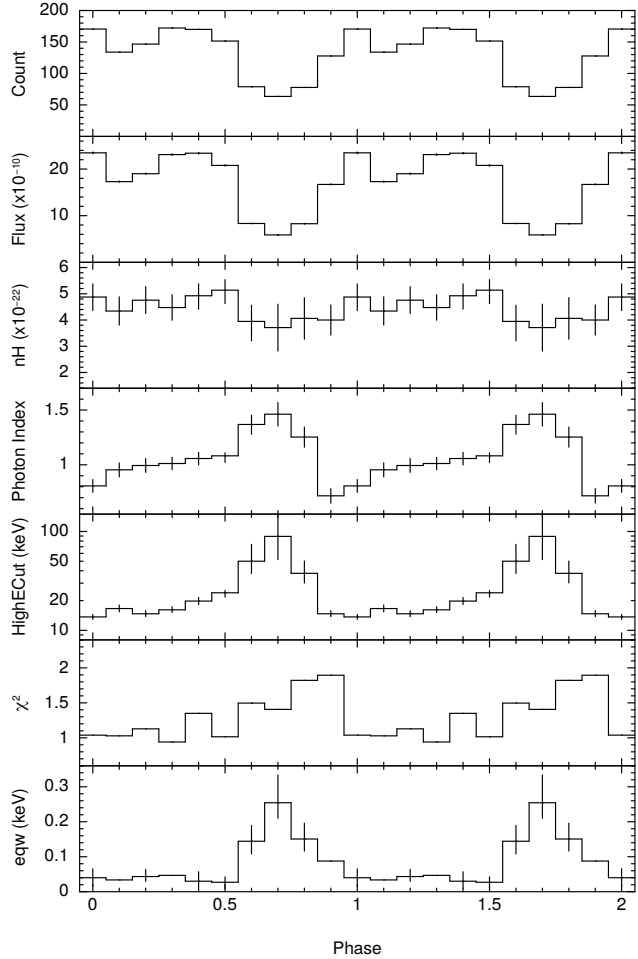}
	\caption{From top to bottom, the panels show the evolution of count rates, flux at 3--25 keV energy band, hydrogen column density, photon index, cutoff energy, reduced $\chi^2$, and equivalent width of the Gaussian component of \textit{RXTE}/PCA observations over phase in MJD 55541--55542. The errors in count rates and flux are too small to be seen in the figure.}
	\label{fig:fasebinparam}
\end{figure}

\begin{figure}
	\centering
	\includegraphics[width=0.5\columnwidth]{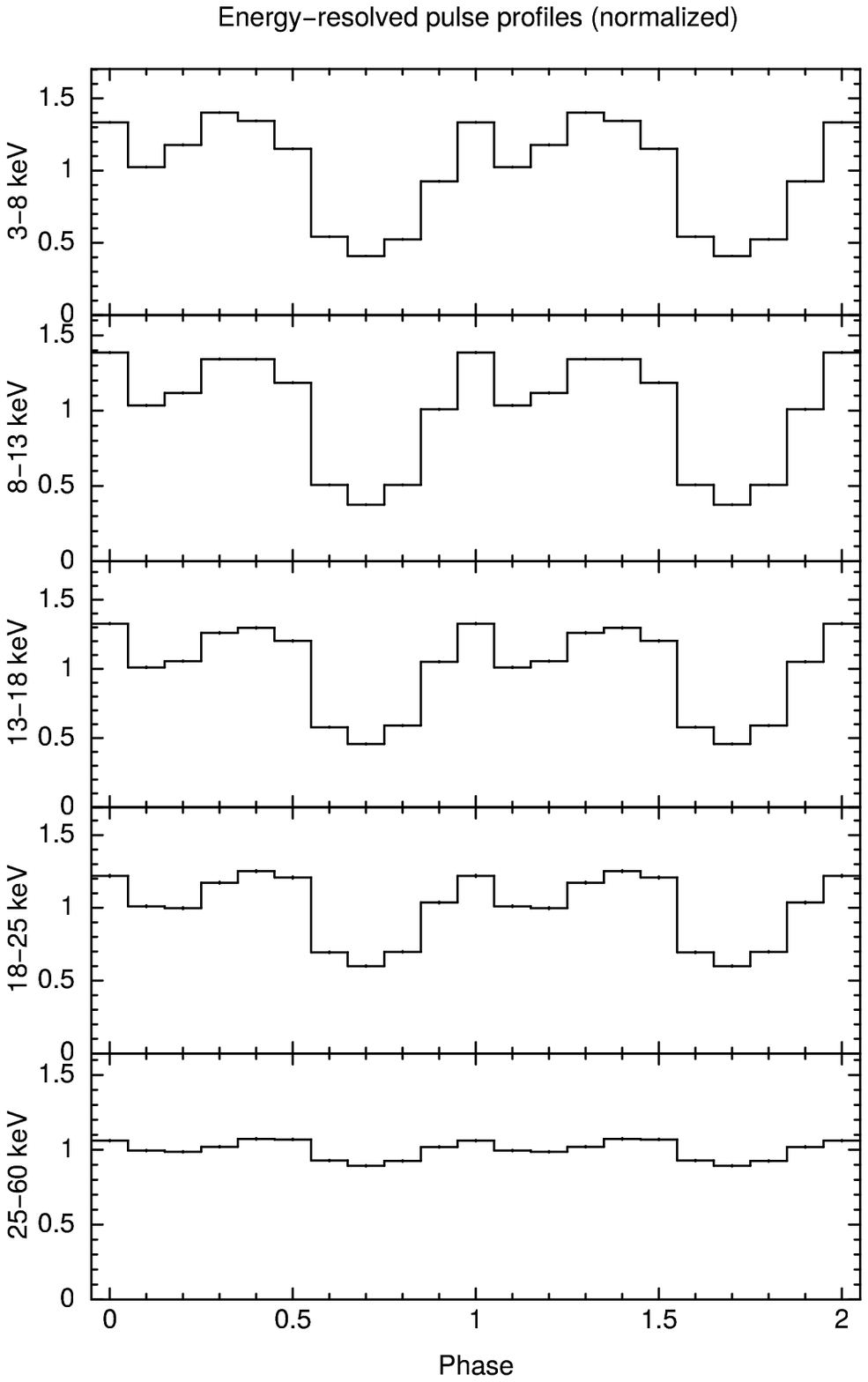}
	\caption{Normalized energy-resolved pulse profiles at various energy ranges of \textit{RXTE}/PCA observations in MJD 55541--55542. The errors in the pulse profiles are too small to be seen.}
	\label{fig:fasebinenergy}
\end{figure}

\section{Summary and Discussion}
\label{sec:discuss}

Using \textit{RXTE}/PCA and \textit{Swift}/XRT observations of MAXI J1409$-$619, we perform timing and X-ray spectral analysis with the aim of constituting a more comprehensive perspective about the nature of this transient source.

After fitting pulse arrival times to a cubic polynomial, we find that the residuals indicate a possible circular orbit with an orbital period 
of 14.7(4) days. This orbital model indicates a mass function of 
about $0.1 M_{\odot}$, implying that the orbital inclination angle should be very small. We also provide an alternative interpretation of these residuals as the noise process due to random torque fluctuations which leads to a torque 
noise strength estimate of $1.3 \times 10^{-18}$ Hz$^2$ s$^{-2}$ Hz$^{-1}$ which is of the 
same order with some wind accreting pulsars in HMXB systems 
and some disc accreting pulsars in LMXB systems.

Torque fluctuations and the noise of accretion powered pulsars have been
studied for several sources \citep{Baykal1993, Bildsten1997}.  
The noise strength associated with torque fluctuations
is obtained for MAXI J1409$-$619 as $1.3 \times 10^{-18}$ Hz$^2$ s$^{-2}$ Hz$^{-1}$. This value of the noise
strength estimate is on the same order with those of other
accretion powered sources such as wind accretors, e.g. Vela X$-$1, 4U 1538$-$52
and GX 301$-$2 with the values changing
between $10^{-20}$ and $10^{-18}$ Hz$^2$ s$^{-2}$ Hz$^{-1}$ \citep{Bildsten1997}. Her
X$-$1 and 4U 1626$-$67, which are disc accretors with
low mass companions, have shown pulse frequency derivatives consistent with
noise strengths
$10^{-21}$ to $10^{-18}$ Hz$^2$ s$^{-2}$ Hz$^{-1}$ \citep{Bildsten1997}. 
Hence, the residuals of the cubic polynomial fit are also consistent with the torque noise fluctuation observed in other accreting sources.

We consider the correlation between pulse frequency derivative and X-ray 
count rate (see Figure \ref{fig:torquelum}) as a consequence of 
disc accretion. Using standard accretion disc theory and assuming the distance to the source as 15 kpc, we estimate the surface dipole magnetic field strength as  
$(2.67\pm 0.44)\times 10^{11}$ Gauss corresponding to a range of inner disc radius value varying between about $0.90\times 10^8$ cm and $1.21\times 10^8$ cm. Our magnetic field
strength estimate is about one order of magnitude smaller than the previous estimate of 
$\simeq3.8\times 10^{12}(1+z)$ Gauss obtained from cyclotron resonance spectral feature 
\citep{Orlandini2012}. Our magnetic field estimate would be consistent with the estimate of \cite{Orlandini2012}, if the distance to the source would be taken as 9.5 kpc. In this case, the inner radius of the accretion disc should also change, ranging between about $5.60\times 10^8$ cm and $7.53\times 10^8$ cm.

QPOs with a very wide range of peak frequencies from 
0.01 Hz to 0.4 Hz have been observed in many accretion powered pulsars 
(see \citet{Acuner2014}, \citet{Inam2004} and references therein). MAXI J1409$-$619
was the first neutron star observed to have harmonics of its QPO feature \citep{Dugair2013}.
Afterwards, similar behaviour is also observed in 4U 0115+63, making MAXI J1409$-$619 and 4U 0115+63 the only two accreting X-ray pulsars known to show QPO harmonics \citep{Roy2019}.

QPOs have usually been interpreted either as the Keplerian frequency at the inner radius of
the accretion disc \citep{vanderKlis87} or the beat frequency of the Keplerian 
frequency at the inner radius of the accretion disc and the spin period of the
neutron star \citep{Alpar85}.

As seen from Figure~\ref{fig:qpofreqflux}, centroid frequencies of detected 
QPOs decrease from 0.2 Hz to 0.1 Hz consistently with decreasing X-ray count rate 
while outburst flux decays. Since these frequencies are much greater than the spin 
frequency of the source, we can solely interpret them to be equal to the
Keplerian frequencies at the inner radius of the accretion disc while the 
inner disc varies with decreasing X-ray flux and then the
inner disc radius can be related to the QPO centroid frequency ($\nu_{QPO}$) as

\begin{equation}
r_0=\left[{{(GM)^{1/2}} \over {2\pi\nu_{QPO}}}\right]^{2/3}.
\label{eqn:QPO}
\end{equation}

From Equation \ref{eqn:QPO}, our QPO frequencies indicate that the inner radius
of the accretion disc moved outward from $r_0=4.9\times 10^8$ cm to 
$r_0=7.8\times 10^8$ cm as the outburst faded. It is important to note that this radius range almost coincides with the radius range (from $5.60\times 10^8$ cm to $7.53\times 10^8$ cm) obtained using standard accretion disc theory if the distance is taken taken as 9.5 kpc. 

In Figure \ref{fig:rxtespectimePIflux}, the spectral fits obtained from the lowest X-ray flux values ($<2\times 10^{-10}$ ergs s$^{-1}$
cm$^{-2}$ corresponding to a 3--25 keV luminosity of about $5\times 10^{36}$ erg s$^{-1}$ or a 2--60 keV luminosity of about $2.4\times 10^{37}$ erg s$^{-1}$) have significantly high power law indices and are modelled by an absorbed power law model without high energy cutoff component. On the other hand, for the measurements plotted as circles in Figure \ref{fig:rxtespectimePIflux}, which are obtained using absorbed power law with  high energy cutoff model, power law index has a weak trend of anti-correlation with X-ray flux up to $\sim10^{-9}$ erg s$^{-1}$ cm$^{-2}$, beyond which it is weakly correlated with X-ray flux. This flux level corresponds to a 3--25 keV luminosity of about $2.5\times 10^{37}$ erg s$^{-1}$ or a 2--60 keV luminosity of about $1.2\times 10^{38}$ ergs s$^{-1}$. This transition level from anti-correlation to correlation can be interpreted as the critical luminosity for which accretion regime shifts from critical (high luminosity) to sub-critical (low luminosity) states. During the transition, the critical luminosity level can be approximately expressed as $\sim1.49\times 10^{37}(B/10^{12}G)^{16/15}$ erg s$^{-1}$ for a neutron star of mass 1.4 M$_{\odot}$ and radius $10^6$ cm  \citep{Becker2012}.
Using this relation regarding critical luminosity, the upper limit for magnetic field strength can be estimated as $\sim7\times 10^{12}$ Gauss for a critical luminosity of $\sim 1.2\times 10^{38}$ ergs s$^{-1}$ (see below for further discussion on the range of critical luminosity), which is closer to the value inferred from cyclotron resonance feature by \cite{Orlandini2012}; however, this value is 26 times greater than the value of magnetic field strength deduced from the torque model described in Section \ref{sect:torquelum}. This discrepancy could be understood (or resolved)  in terms of the torque  model in the following way: Since the angular acceleration is proportional to both magnetic moment and accretion luminosity $\dot{\nu}\sim \mu^{2/7} L^{6/7}$, the disagreement of the magnetic field values can be compensated if the luminosity is about $\sim34 \%$ of its original value corresponding to a distance of about 8.7 kpc.

Analogous to the behaviour of MAXI 1409-619, transition from anticorrelation to correlation between the photon index and the X-ray luminosity is observed in EXO 2030+375 \citep{Epili2017}. Positive and negative correlations with power law index are also seen in sources with cyclotron features \citep{Klochkov2011,Malacaria2015}. For MAXI J1409-619,  it should be noted that the starting flux level of positive correlation of spectral index may represent the upper limit of critical luminosity of the source, since the spectral index values are scattered and obtained via diferent spectral models at lower flux values.\footnote{ At very low flux levels (see Figure \ref{fig:rxtespectimePIflux}), high energy cutoff power law model could not be applied due to the deteriorated spectral quality; therefore, only absorbed power law model is used. As two different spectral models are introduced, the spectral parameters at lower flux values ($\lesssim5\times10^{-10}$ erg s$^{-1}$ cm$^2$ or $6\times 10^{37}$ erg s$^{-1}$) should be examined with caution.}

 On the other hand, it is important to note that the pulse fraction of the source is correlated with the RXTE/PCA count rate up to 200 cts s$^{-1}$ (see Figure \ref{fig:pulsedfrac}),
 corresponding to a 2--60 keV luminosity of about $\sim 6 \times 10^{37}$ ergs s$^{-1}$ if a distance of 15 kpc is assumed. Beyond this luminosity level, the pulsed fraction anticorrelates with the count rate. The observed anticorrelation can be interpreted as an increase of unpulsed component of the emission of the pulsar in the form of fan beam, as suggested by \cite{Becker2012} (see also \cite{Mushtukov2015}). Similar kind of mode switching of correlation around critical luminosity is also observed in 2S 1417-624; at lower flux level \cite{Inam2004b} reported a positive correlation of pulse fraction and flux, while \cite{Gupta2019} reported a negative correlation for the same source when the flux level is higher. Transition from anticorrelation to correlation
between the pulsed fraction and the X-ray luminosity is also observed in Swift J0243.6+6124 for which the flux level of the transition is interpreted as the critical luminosity \citep{wilsonhodge2018}.
 MAXI 1409-619 is a remarkable source that shows correlation mode switching of both power law index and pulse fraction around the critical luminosity. Utilising those features, the critical luminosity level of MAXI J1409-619 can be inffered within the range of $6\times 10^{37}$ erg s$^{-1}$ to $1.2\times 10^{38}$ ergs s$^{-1}$, where the upper limit is deduced from the starting flux level of positive photon index correlation and the lower limit is derived from the flux level beyond which the pulsed fraction anticorrelates with the count rate. Using the relation suggested by \cite{Becker2012}, the inferred critical luminosity range corresponds to magnetic field range of $(3.7-7)\times 10^{12}$ Gauss which might imply that the actual source luminosity is between 34$\%$ to 41$\%$ of the luminosity deduced from 15 kpc distance estimation, corresponding to a distance range between 8.7 kpc to 9.6kpc. Although this distance range does not coincide with the most likely distance estimate reported as 15 kpc, \citet{Orlandini2012} did not rule out the possibility of a source distance of $\sim 7.9$ kpc which is more compatible with this range. 

From the phase-resolved spectrum of the source presented in Figure \ref{fig:fasebinparam},
we find that the neutral hydrogen column density $n_H$ does not show a phase dependence whereas power law index and equivalent width of the iron line tend to have an anti-correlation with the source flux. 
Higher photon index values at the minimum flux can be explained by hard emission of 
the polar region. As the pulsed flux diminishes, the observed emission is 
softened. At the pulse minima, the high energy cutoff value goes out of spectral fitting range (3--25 
keV for RXTE observations).  This is probably the result of the decreased signal-to-noise ratio 
due to the decreased count rate of the source. Besides, the phase bins 
corresponding to the pulse minima have larger $\chi^2$ values, possibly 
owing to the excess high-energy cutoff values obtained by the fit and the 
smaller signal-to-noise ratios as explained above (see Fig. \ref{fig:fasebinspecbestworst} for examples of low and high $\chi^2$ spectra). A possible explanation of the increased equivalent width of the Gaussian iron K$\alpha$ line at pulse minimum is the arise of 7.04 keV K$\beta$ line emission as the emission from the hotspot fades from view. Along with the 6.4 keV K$\alpha$ line emission, blending of two iron lines results in Gaussian line broadening \citep{Serim2017}.
The broadening of the iron line at lower flux might be explained by the emergence of K$\beta$ emission when the enhanced emission of the hotspot fades away. 
Anti-correlation of the power law index and flux implies that the pulsed emission is relatively hard when compared to the pulse minima. 

\section*{Acknowledgements}

We acknowledge support from T\"{U}B\.{I}TAK, the Scientific and 
Technological Research Council of Turkey through the research project MFAG 118F037. 
AB acknowledges MPE and Werner Becker for helpful discussions. We
would like to express our gratitudes to the anonymous referee for the valuable remarks that
ease the improvement of the manuscript.




\bibliographystyle{mnras}
\bibliography{bib} 








\bsp	
\label{lastpage}
\end{document}